\documentclass[12pt]{revtex4}
\usepackage{latexsym, color, graphicx, comment}
\usepackage{amssymb}
\usepackage{amsmath}

\newcommand{\vect}[1]{\mbox{\boldmath $#1$}}
\newcommand{\erf}{\mathrm{erf}}
\newcommand{\PS}{Pfirsch-Schl\"{u}ter~}
\newcommand{\ExB}{\vect{E}\times\vect{B}}
\newcommand{\Cnl}{C_{\mathrm{nl}a}}
\newcommand{\vma}{\vect{v}_{\mathrm{m}a}}
\newcommand{\vm}{\vect{v}_{\mathrm{m}}}
\newcommand{\vda}{\vect{v}_{\mathrm{d}a}}
\newcommand{\vE}{\vect{v}_{E}}
\newcommand{\vdao}{\vect{v}_{\mathrm{d}a0}}
\newcommand{\vdo}{\vect{v}_{\mathrm{d}0}}
\newcommand{\vEo}{\vect{v}_{E0}}
\newcommand{\vpar}{v_{||}}
\newcommand{\vperp}{v_{\bot}}
\newcommand{\rhop}{\rho_{\theta}}
\newcommand{\rT}{r_{T}}
\newcommand{\Ti}{T_{\mathrm{i}}}
\newcommand{\rTi}{r_{T\mathrm{i}}}

\newcommand{\rn}{r_{n}}
\newcommand{\rperp}{r_\bot}
\newcommand{\fM}{f_{\mathrm{M}}}
\newcommand{\fMa}{f_{\mathrm{M}a}}
\newcommand{\fMi}{f_{\mathrm{M}i}}

\newcommand{\nuPrime}{\hat{\nu}}

\newcommand{\psia}{\psi_0}
\newcommand{\psiaHat}{\hat{\psi}_0}
\newcommand{\changed}[1]{#1}

\begin{document}

\title{Radially global $\delta f$ computation of neoclassical phenomena in a tokamak pedestal}



\author{Matt Landreman}
\email[]{mattland@umd.edu}
%\thanks{}
\affiliation{Institute for Research in Electronics and Applied Physics, University of Maryland, College Park, MD, 20742, USA}
%\author{co-authors}
\author{Felix I Parra}
\affiliation{Department of Physics, University of Oxford, Oxford, OX1 3PU, UK}
\author{Peter J Catto}
\author{Darin R Ernst}
\author{Istvan Pusztai}
\affiliation{Plasma Science and Fusion Center, MIT, Cambridge, MA, 02139, USA}


\date{\today}

\begin{abstract}

Conventional radially-local
neoclassical calculations become inadequate if
the radial gradient scale lengths of the H-mode pedestal become as small as the poloidal ion gyroradius.
Here, we describe a radially global
$\delta f$ continuum code that generalizes
neoclassical calculations to allow stronger gradients.
As with conventional neoclassical
calculations,
the formulation is time-independent and requires only the solution of a single sparse linear system.
We demonstrate precise agreement with an asymptotic analytic solution
of the radially global kinetic equation
in the appropriate limits of aspect ratio and collisionality.
This agreement depends crucially on accurate treatment of finite orbit width effects.

\end{abstract}


\maketitle %\maketitle must follow title, authors, abstract and \pacs

\section{Introduction}

Neoclassical effects\cite{HintonHazeltine, PerBook} are likely to be important
in the pedestal of an H-mode tokamak for several reasons.
First, neoclassical flow, current, and radial fluxes will be large in the pedestal
as they are driven by gradients, which are large in the region.
Second, turbulence must be somewhat suppressed in the pedestal
in order for the gradients to become so large,
so the collisional radial ion heat flux may
be a large fraction of the total ion energy transport.
Third, the bootstrap current may affect edge stability.
However, neoclassical effects are usually calculated using an ordering
which can be invalid in the pedestal.
Specifically, conventional neoclassical calculations
assume $\rhop/\rperp \ll 1$, where $\rhop$ is the poloidal ion gyroradius,
and $\rperp$ is the scale length for the pressure or temperature.
Empirically, $\rhop/\rperp$ can be comparable to 1 in the pedestal
due to the strong gradients,
in which case conventional neoclassical calculations are inadequate.
Physically, $\rhop$ is (up to a geometric factor) the width
of drift orbits, so conventional neoclassical calculations assume the orbit
width is thin compared to equilibrium scales, whereas finite-orbit-width effects
may become significant in the pedestal.

For improved pedestal neoclassical calculations, some
researchers\cite{Lin1,Lin2,Wang1,CSChang1,Wang2,CSChang2,Wang3,Kolesnikov,ORB5, Xu1,Xu2,XGCBootstrap,Dorf}
have considered a model that may be termed ``global full-$f$'',
in contrast to the conventional approach\cite{HintonHazeltine, PerBook, WongChan, NEOFP, JeffParker} which may be called ``local $\delta f$.''
In a full-$f$ code, the entire distribution function $f$ is solved for,
whereas in a $\delta f$ code, one only solves for the departure from a Maxwellian,
with the Maxwellian constant on each flux surface.  Due to both the
nonlinearity of the Fokker-Planck collision operator
and to an $\ExB$ nonlinearity, the full-$f$ problem is nonlinear
in the unknowns, whereas a $\delta f$ problem is linear in the unknowns.
The preceding distinction of local vs. global refers to whether radial
derivatives of the unknown distribution function are retained in the kinetic equation.
The radial coordinate is only a parameter in a local code,
whereas it is more challenging to solve a global problem due to radial coupling.
Global full-$f$ numerical calculations therefore require orders of magnitude more
processor-hours\cite{XGCBootstrap} than local $\delta f$ calculations.

An intermediate ``global $\delta f$'' model represents a happy medium in some circumstances.
In the global $\delta f$ approach, an expansion about a Maxwellian flux function is still made so
the problem remains linear, but radial derivatives of the unknown distribution are retained.
Therefore, the global $\delta f$ approach formally allows stronger radial gradients than
the local $\delta f$ approach. However, the global $\delta f$ approach does not
allow as strong an ion temperature gradient as the global full-$f$ approach
because the departure from a Maxwellian flux function is driven by
the ion temperature gradient.
The density gradient does not drive a departure from a Maxwellian if
the ions are electrostatically confined.
Detailed analysis\cite{Grisha1, usPedestal} shows the global $\delta f$
model requires $r_{Ti} \gg \rhop$ and $r_{\eta i} \gg \rhop$ where
$\eta_i = n \exp(e \Phi/T_i)$, whereas $r_n$ is allowed to be as small as $\rhop$ or smaller.
Here, $r_X$ denotes the radial scale length of $X$, $n$ is the density, $e$ is the proton charge, $\Phi$ is the electrostatic potential,
and $T_i$ is the ion temperature.
The local $\delta f$ model also requires not only $r_{Ti} \gg \rhop$ but also $r_n \gg \rhop$.
(All three models allow the electron temperature scale length $r_{Te}$ to be as small as $\rhop$.)
Thus, the global $\delta f$ model
is more general than the local $\delta f$ model but not as general as global full-$f$ model.
The global $\delta f$ model allows strong density pedestals with weaker
ion temperature gradient,
and some experiments indeed show this situation in the
pedestal\cite{GroebnerOsborne, Maggi, Corre, Groebner, Morgan, Diallo, Meyer, Sontag, SauterProfiles, Maingi, Putterich}.
Both entropy considerations \cite{Grisha1} and data \cite{Morgan, Meyer} suggest $r_{Ti}$ resists  becoming as small as $\rhop$ when collisionality is low, while $r_n$ is not similarly constrained,
supporting the global $\delta f$ ordering.
It is also useful to study the global $\delta f$ approach because it incorporates
some but not all of the elements required for a full-$f$ code.
For example, the linear system solved in a global $\delta f$ code closely resembles the Jacobian system
that must be solved in a Newton-like iteration for a nonlinear full-$f$ code,
but only a single solve is needed rather than many.
A full-$f$ code will necessarily be more complicated than a $\delta f$ code, so the latter can be useful
for benchmarking the former.
The relationship between local and global $\delta f$ neoclassical codes
is quite analogous to the relationship between flux-tube and global $\delta f$ gyrokinetic codes.

In this work we detail the implementation of the global $\delta f$ model in a new code
PERFECT (Pedestal and Edge Radially-global Fokker-Planck Evaluation of Collisional Transport.)
An arbitrary number of species may be included,
and parallel iterative Krylov-space solvers are implemented for efficiency.
Very different algorithms are used compared to previous codes such as Refs. \cite{usPedestal, Dorf},
and in particular, a time-independent rather than time-dependent approach is used,
so the solver convergence is not limited by the timescale of physical relaxation.
Nested flux surface geometry of arbitrary shaping is allowed, as is arbitrary collisionality,
and the full linearized Fokker-Planck
collision operator is implemented with arbitrary mass ratio and charge permitted.

Several issues must be addressed in a global $\delta f$ code
which do not arise in a local $\delta f$ code. One issue is the need for boundary conditions
in the radial coordinate. Another issue is the necessity of sources and/or sinks in the kinetic
equation. As shown in Ref. \cite{usPedestal}, for general input profiles, the global $\delta f$ kinetic
equation has no time-independent solution unless a sink term is present.
Without this term, the departure from the Maxwellian would grow in time,
eventually violating the linearization.
An analogous issue arises in global $\delta f$ gyrokinetic codes \cite{GENESource}.
Physically, given the input profiles of density, temperature, and radial electric field,
radial neoclassical fluxes arise which generally have some divergence, acting to alter
the input profiles. In a local calculation, the timescale of this profile change is formally separated
from the timescale considered in the kinetic equation, but no such timescale separation is formally imposed
in the global $\delta f$
model due to the strong radial gradients allowed.
In a real plasma, the divergence of the neoclassical fluxes
may be canceled by a divergence in the turbulent fluxes,
but in a purely neoclassical calculation,
some other term must be included to balance the divergence.
The implementation of the radial boundary conditions and sink in PERFECT
will be detailed in section \ref{sec:numerical}.

Several authors have investigated the global $\delta f$ model analytically \cite{Grisha1, GrishaNeo, GrishaNeoErratum, GrishaPRL, Istvan, IstvanErratum, Peter}.
However, little work has been done to
compare the departures from conventional neoclassical results predicted by
the aforementioned first-principles theories to direct numerical solution of the kinetic equation.
To our knowledge,
the only published comparison of this kind to date is
in Ref. \cite{Dorf},
in which some qualitative agreement was seen, but
not precise quantitative agreement.
One purpose of the work here is to demonstrate precise quantitative
agreement between theory and simulation.
Due to the complexity of edge plasma transport codes,
it is crucial to have problems on which they can be benchmarked
to ensure the codes are free of errors.
Our work illustrates such a problem in which the effects of finite orbit width
are central.

\changed{
As with solutions of the conventional neoclassical kinetic equation, in our procedure we compute the response of particles
to prescribed electromagnetic fields. As such, the calculation results in one relation between the radial electric field $E_r$ and the flow
(i.e. (\ref{eq17})).  However, with only one relation between these two quantities, neither quantity is truly determined
unless an additional constraint is introduced. The relevant additional condition would be a momentum transport relation.
Momentum transport in tokamaks is a complicated problem, involving turbulence \cite{Felix2010}, and so we do not attempt to model it in the present work.
We focus here on the first relation, determining the flow as a function of a prescribed $E_r$ profile.
If in future work the calculation here were coupled to a turbulence code, it may be possible to obtain a self-consistent $E_r$,
thereby eliminating the need to specify $E_r$ as an input.
}

In the next section, we more precisely define the global full-$f$, global $\delta f$, and local $\delta f$ models.
We then describe the numerical implementation of the global $\delta f$ model
in section \ref{sec:numerical}.
In section \ref{sec:plateau}, we summarize the analytic theory to which
the numerical results are compared.
The comparisons between the code and analytic theory are
then detailed in section \ref{sec:comparison}.
Simulation results for several other conditions are shown in
section \ref{sec:advanced},
and we discuss the results and conclude in section \ref{sec:conclusions}.

\section{Definitions and physics model}
\label{sec:model}

We next detail the three versions of the drift-kinetic equation (global full-$f$,
global $\delta f$, and local $\delta f$), highlighting the assumptions made in each simplification.
Throughout, time derivatives will be neglected, since our focus
is the neoclassical equilibrium.

We begin with the global full-$f$ drift-kinetic equation:
\begin{equation}
\left( \vpar\vect{b} + \vda\right) \cdot (\nabla f_a)_{\mu,W_a} = \Cnl + S_a
\label{eq:fullf}
\end{equation}
where $a$ denotes particle species, $f_a$ is the total distribution function, $\vect{b} = \vect{B}/B$, $B = |\vect{B}|$,
$ \vda =(\vpar/\Omega_a)\left( \nabla
\right)_{\mu ,W_a} \times \left( {v_{\vert \vert } \vect{b}} \right)
=\vE+\vma$,
$\vE=cB^{-2}\vect{B}\times\nabla \Phi$,
\begin{equation}
\vma = \frac{m_a cv_{\vert \vert
}^2 }{Z_a eB}\nabla \times \vect{b}+\frac{m_a cv_\bot^2
}{2Z_a eB^3}\vect{B}\times \nabla B,
\label{eq:drift}
\end{equation}
$\Phi$ is the electrostatic potential,
$\Cnl$ is the nonlinear
Fokker-Planck collision operator, and $S_a$ represents any sources and sinks.
Subscripts on partial derivatives indicate
quantities held fixed, which here are $\mu=\vperp^2/(2B)$ and total energy
$W_a = v^2/2+Z_a e\Phi/m_a$.
As noted in Ref \cite{Hazeltine}, (\ref{eq:fullf}) may be derived recursively in a manner that does
not require $\vda\cdot(\nabla f)_{\mu, W_a} \ll \vpar \vect{b}\cdot(\nabla f)_{\mu, W_a}$.
Indeed, we will eventually allow some radial scale lengths to be comparable to $\rhop$ (ordering $\rhop \gg \rho$ where $\rho$ is the gyroradius) so these two terms on the left-hand side of (\ref{eq:fullf}) can be comparable.
The form of the drifts (\ref{eq:drift}) is conservative in that moments of (\ref{eq:fullf})
give the desired conservation laws for mass, momentum, and energy, as shown in appendix B of Ref. \cite{usPedestal}.
In (\ref{eq:drift}), notice $\nabla\times\vect{b} = \vect{b}\times(\vect{b}\cdot\nabla\vect{b}) + \vect{b}\vect{b}\cdot\nabla\times\vect{b}$
contains the curvature drift and a parallel drift.
The latter is altered \cite{Boozer80,Parra} when higher order corrections to the drifts and magnetic moment
are retained, but these corrections are unimportant for our purposes.

Even if the flux-surface-averaged densities and temperatures of each species are specified along with the magnetic field
and flux-surface-averaged potential, (\ref{eq:fullf}) is nonlinear in the unknowns ($f_a$ and the poloidally varying part of $\Phi$)
both due to the $\vE\cdot\nabla f_a$ nonlinearity, and also due to the bilinearity of the collision operator.
Even if $\Cnl$ is replaced with a ``linearized'' collision operator $\propto n_a f_a$ where $n_a = \int d^3 v\, f_a$ is the density,
the collision term is still nonlinear in the unknowns if the poloidal variation of $n_a$ is considered unknown prior to solution of (\ref{eq:fullf}).

We next proceed to linearize the problem.
To do so, we must require that the
potential $\Phi $ be constant on flux surfaces to leading order, and we will
later refine the conditions under which this assumption is justified.
We let
$\Phi_1=\Phi -\Phi_0 $
where $\Phi_0 =\left\langle \Phi \right\rangle $.
Here, $\left\langle \right\rangle $ denotes a flux surface average,
which for any
quantity $X$ is
\begin{equation}
\label{eq24}
\left\langle X \right\rangle =\frac{1}{{V}'}\int_0^{2\pi }
{\frac{X\mbox{\thinspace }d\theta }{\vect{B}\cdot \nabla \theta }},
\end{equation}
where
${V}'=\int_0^{2\pi } d\theta / \vect{B}\cdot \nabla \theta $.
We also define the leading-order total energy
$W_{a0} =v^2/2+ Z_ae\Phi_0 /m_a$,
which will be used as an independent variable for the rest of this section,
and we define the drift
$\vdao=\vEo+\vma$ where $\vEo=cB^{-2}\vect{B}\times\nabla\Phi_0$.

We take the distribution function of each species to be approximately the Maxwellian
\begin{equation}
\label{eq6}
\fMa
 =\eta_a \left( \psi \right)\left[ {\frac{m_a}{2\pi T_a\left( \psi \right)}}
\right]^{3/2}\exp \left( {-\frac{m_a W_{a0} }{T_a\left( \psi \right)}} \right),
\end{equation}
where $2\pi \psi $ denotes the poloidal flux.
The leading-order density $n_a\left( \psi \right)$ and temperature $T_a\left(
\psi \right)$ are flux functions, as is the
``pseudo-density''
$\eta_a \left( \psi \right)=n_a\exp \left( Z_a e\Phi_0 /T_a \right)$.
To obtain a linear kinetic equation, we must assume $Z_a e \Phi_1/T_a \ll 1$, and soon we will examine how well this inequality is satisfied.  We order the species charges $Z_a \sim 1$.
Next, the ``nonadiabatic" part of the distribution function $g_a$ is defined by
\begin{equation}
\label{eq9}
g_a=f_a-\fMa +\frac{Z_a e\Phi_1 }{T_a}\fMa.
\end{equation}
We require that
$g_a/\fMa \ll 1$
so $f_a \approx \fMa$ and so the collision operators may be linearized about $\fMa$.
The small ratios $g_a/\fMa$ and $e\Phi_1/T_a$ are linked through quasi-neutrality.
For example, consider a pure plasma,
in which electrons and ions may be denoted with subscripts $e$ and $i$.
Noting that $g_e \ll g_i$ since $g_a$ scales in conventional neoclassical theory as $\sim \sqrt{m_a}$,
quasi-neutrality gives
\begin{equation}
\label{eq13}
\frac{e\Phi_1 }{T_i}=\left( {\frac{T_i}{T_e }+Z_i} \right)^{-1}\frac{1}{n_i}\int
{d^3v \mbox{\thinspace }g_i}.
\end{equation}
Thus, $e\Phi_1/T_i \sim g_i/\fMi$.

Substituting (\ref{eq9}) into the full-$f$ kinetic equation (\ref{eq:fullf}),
and using $e\Phi_1/T_a \ll 1$, we arrive at the global $\delta f$ equation:
\begin{equation}
\label{eq15}
\left( {v_{\vert \vert } \vect{b}+\vdao } \right)\cdot
\left( {\nabla g_a} \right)_{\mu ,W_{a0} } -C_{\ell a} \left\{ g_a \right\}-S_a
=-\vma\cdot\nabla\psi \left(\frac{\partial \fMa}{\partial\psi}\right)_{W_{a0}}
\end{equation}
where $C_{\ell a} $ is the collision operator linearized about the Maxwellians
(\ref{eq6}).
To obtain $\vda\cdot\nabla\psi \approx \vma\cdot\nabla\psi$ in (\ref{eq15}),
we have noted $\vEo\cdot\nabla\psi=0$ and $(\vE\cdot\nabla\psi) / (\vma\cdot\nabla\psi) \sim Z_a e \Phi_1 / (\epsilon T_a)$
and taken this ratio to be small. Here,
$\epsilon$ is the inverse aspect ratio, and we will not treat $\epsilon$ as an expansion parameter except in section \ref{sec:plateau}.
To neglect $(c/B^2) \vect{B}\times\nabla\Phi_1\cdot\nabla\theta$ in obtaining (\ref{eq15}),
we have used $\partial\Phi_1/\partial\psi \ll d\Phi_0/d\psi$.
All terms involving $\Phi_1 $ have been rigorously accounted for
($\ExB$ drift and parallel acceleration), and
it can be seen that all $\Phi_1 $ dependence has disappeared from (\ref{eq15}),
since it has either been absorbed into the Boltzmann response in (\ref{eq9}) or because it is formally negligible. Thus,
(\ref{eq15}) is linear in the unknowns $g_a$.

Since (\ref{eq15}) is a linear inhomogeneous equation, the size of the solution
$g_a$ is linear in the size of the inhomogeneous term on the right hand side,
which is in turn proportional to
\begin{equation}
\left(\frac{\partial \fMa}{\partial\psi}\right)_{W_{a0}}
= \left[ \frac{1}{\eta_a} \frac{d\eta_a}{d\psi} + \left( \frac{m_a W_{0a}}{T_a} - \frac{3}{2}\right) \frac{1}{T_a}\frac{d T_a}{d\psi}\right]
\fMa
\label{eq:gradfM}
\end{equation}
If the radial gradients of $\eta_a$ or $T_a$ become sufficiently large, then, $g_a$ will become as large as $\fMa$, violating the ordering.
The gradient at which this transition occurs may be estimated by balancing $\vpar \vect{b}\cdot(\nabla g_a)_{\mu, W_{a0}}$ (the first
term of of (\ref{eq15})) with the right-hand side
of (\ref{eq15}), leading to the requirements
\begin{equation}
r_{\eta a} \gg \rho_{\theta a} \;\;\;\mbox{and}\;\;\; r_{Ta} \gg \rho_{\theta a}.
\label{eq:weakGradients}
\end{equation}
Here,
$r_X =\left| X /\left( |\nabla\psi| dX/d\psi \right)  \right|$
for any $X$,
$\rho_{\theta a}=v_a m_a c/\left( {Z_a e B_\theta } \right)$ is the poloidal gyroradius,
$B_\theta $ is the poloidal magnetic field, and $v_a =\sqrt {2T_a/m_a} $.
Due to the scaling $\rho_{\theta a} \sim \sqrt{m_a}/Z_a$,
(\ref{eq:weakGradients}) is typically harder to satisfy for the main ions than for impurities or electrons.
However, notice that gradients of $n_a$ and $\Phi_0$ do not appear in (\ref{eq:gradfM}),
so these gradients do not drive departure from a Maxwellian.
\changed{Therefore, $r_{na}$ and $r_{\Phi_0}$ need not be $\gg \rho_{\theta a}$ in
the global $\delta f$ model. The minimum allowed length for these two gradient scale lengths
is rather the less restrictive limit $\rho_a$, since (\ref{eq:fullf})
is valid when the $\vect{E}\times\vect{B}$ flow is subsonic and when $f_a$ does not vary
significantly on the $\rho_a$ scale.}
%Therefore there is no limit on these gradients
%in the global $\delta f$ model, as long as the inequalities in (\ref{eq:weakGradients}) remain satisfied %\cite{Grisha1}.
Indeed, we will take $r_{ni}\sim \rho_{\theta i}$ in the pedestal calculations in later sections.
The situation $r_{na} \ll r_{\eta a}, r_{Ta}$ corresponds to electrostatic ion confinement.
Physically, for a given electrostatic potential, electrostatic confinement is a situation
of thermodynamic equilibrium. Thus, a steep density gradient does not drive $g_a$ (the departure from thermodynamic equilibrium)
as long as the confinement is nearly electrostatic.

We may now re-examine our earlier approximation $Z_a e \Phi_1/(\epsilon T_a) \ll 1.$
Noting (\ref{eq13}), and again balancing the first and last terms of (\ref{eq15}) using (\ref{eq:gradfM}),
we estimate $Z_a e \Phi_1/(\epsilon T_a) \sim \max(\rho_{\theta a}/r_{\eta a}, \rho_{\theta a}/r_{Ta})$.
As we have already had to assume (\ref{eq:weakGradients}), our assumption on the size of $\Phi_1$
is therefore self-consistent and is not an extra restriction.
As a practical matter, given any input profiles $T_a(\psi)$, $n_a(\psi)$, and $\Phi_0(\psi)$,
(\ref{eq15}) may be solved for $g_a$, and as long as
$\int d^3 v\, g_a \ll n_a$ and $\int d^3v\, \vpar g_a \ll n_a v_a$,
the linearization leading to (\ref{eq15}) is reasonable.

It is useful to order $B_\theta /B \ll 1$, which is reasonable for the edge of conventional tokamaks.
This ordering ensures the distinction between guiding-center position and particle position is unimportant,
even when $n_i$ varies on the $\rho_{\theta i}$ scale
since $\rho_{\theta i} \gg \rho_{i} = v_i/\Omega_i$ with $\Omega_i = e B/(m_i c)$.

Although we require $g_a \ll \fMa$, we allow $(\partial g_a/\partial \psi)_{\mu,W_{a0}}$ to compete with $(\partial \fMa/\partial\psi)_{\mu, W_{a0}}$
in (\ref{eq15}). These terms may be comparable
because (\ref{eq:weakGradients}) ensures the radial scale length of $\fMa$ is $\gg \rho_{\theta a}$,
whereas when $r_{na}\sim\rho_{\theta a}$,
the radial scale length of $g_a$ may be as small as $r_{na}$.

The local $\delta f$ drift-kinetic equation is obtained by neglecting the $\vdao\cdot(\nabla g_a)_{\mu, W_{a0}}$
term in (\ref{eq15}):
\begin{equation}
\label{eq:local}
 v_{\vert \vert } \vect{b}\cdot
\left( {\nabla g_a} \right)_{\mu ,W_{a0} } -C_{\ell a} \left\{ g_a \right\}-S_a
=-\vma\cdot\nabla\psi \left(\frac{\partial \fMa}{\partial\psi}\right)_{W_{a0}}.
\end{equation}
(Note that we take ``local'' to mean not only that radial drifts acting on $g_a$ are neglected,
but also that poloidal drifts are neglected.)
In the absence of sources,
(\ref{eq:local}) is the equation solved in conventional neoclassical theory \cite{HintonHazeltine, PerBook} and codes \cite{WongChan, NEOFP}.
Comparing the $\vdao\cdot(\nabla g)_{\mu, W_{a0}}$ term neglected in this local model
to $\vpar \vect{b}\cdot(\nabla g_a)_{\mu, W_{a0}}$,
and recalling the radial variation of $g_a$ is $\sim r_{na}$,
then the local model requires $r_{na} \gg \rho_{\theta a}$,
which was not required for the global $\delta f$ model.
Furthermore, $(\vEo\cdot\nabla\theta)/(\vpar \vect{b}\cdot\nabla\theta)\sim \rho_{\theta a} / r_{\Phi}$,
so neglect of the poloidal drift in the local model also requires
$r_{\Phi} \gg \rho_{\theta a}$,
which was not required for the global $\delta f$ model.
To understand these two new restrictions
from another perspective, consider that the characteristic curves of the local $\delta f$ equation are on a constant-$\psi$ surface
-- infinitely thin compared to equilibrium scales -- whereas the characteristic curves of the global $\delta f$
equation have finite width. Physically, then, the local $\delta f$ model assumes no equilibrium quantities may vary over the orbit width $\sim \rho_{\theta a}$,
whereas the global $\delta f$ model does allow variation in density and potential over the orbit width.

Analytic solution of the kinetic equations (\ref{eq15}) or (\ref{eq:local})
requires additional subsidiary expansion in collisionality.
The ranges of collisionality may be defined using
$\nuPrime=\nu_{aa} q R/v_a$
and
$\nu_*=\nuPrime/\epsilon^{3/2}$,
where
$\nu_{aa} =4\sqrt {2\pi } Z_a^4 e^4 n_a\ln \Lambda/(3\sqrt m_a T_a^{3/2})$
is the self-collision frequency,
$\ln\Lambda$ is the Coulomb logarithm,
$q$ is the safety factor, and $R$ is the major radius.
The \PS (collisional) regime is $1 \ll \nuPrime$,
the plateau regime is defined by $\nuPrime \ll 1 \ll \nu_*$,
and the banana (low collisionality) regime
is defined by $\nu_* \ll 1$.

Once $g_a$ is obtained analytically or numerically, several interesting moments of the total distribution
function can be computed. These moments include the parallel flow
\begin{equation}
\label{eq17}
V_{a\vert \vert } =\frac{1}{n_a}\int {d^3v \mbox{\thinspace }} v
_{\vert \vert } g_a,
\end{equation}
the radial particle flux
\begin{equation}
\label{eq21}
\left\langle {\int {d^3v \mbox{\thinspace }} g_a\vma \cdot
\nabla \psi } \right\rangle ,
\end{equation}
the radial momentum flux
\begin{equation}
\label{eq22}
\left\langle {\int {d^3v \mbox{\thinspace }} g_a\frac{Iv_{\vert
\vert } }{B}\vma \cdot \nabla \psi } \right\rangle ,
\end{equation}
and the radial energy flux
\begin{equation}
\label{eq23}
\left\langle {\int {d^3v \mbox{\thinspace }} g_a\frac{m_a v
^2}{2}\vma \cdot \nabla \psi } \right\rangle .
\end{equation}
Here, $I(\psi)=RB_\zeta$ is the major radius times the toroidal field.
The three radial fluxes arise in conservation
equations, a derivation of which can be found in Appendix B of Ref. \cite{usPedestal}.
For subsonic ions in a plasma with no non-trace impurities, the flow is known
to have the form
\begin{equation}
\label{eq17}
V_{i\vert \vert } =-\frac{cI}{Z_i eB}\left( {\frac{1}{n_i}\frac{dp_i}{d\psi
}+Z_i e\frac{d\Phi_0 }{d\psi }-k \frac{B^2}{\left\langle {B^2}
\right\rangle }\frac{dT_i}{d\psi }} \right)
\end{equation}
where $k$ is a dimensionless coefficient.
In the local limit,
the flow coefficient $k$ is constant on each flux surface.
However, in the radially global case considered hereafter,
$k$ may vary\cite{usPedestal} with $\theta$.

\section{Boundary conditions, sources, and numerical solution}
\label{sec:numerical}

We now detail the procedure for solving the global $\delta f$ kinetic equation (\ref{eq15})
implemented in PERFECT, which is improved compared to the approach of Ref. \cite{usPedestal} in several respects.
Instead of solving a time-dependent problem to dynamically determine the equilibrium,
here we directly solve the time-independent system, which is substantially faster.
We also employ a new treatment of the volumetric sources and sinks, described later in this section,
and improved radial boundary conditions.

\begin{figure}[h!]
% Original figure generated by m20130113_01_morpheusBoundaryConditionFigure.m
% Later modified in PowerPoint and printed to PDF.
%\includegraphics[width=3.5in]{PERFECTBoundaryConditions.pdf}
\includegraphics[width=3.5in]{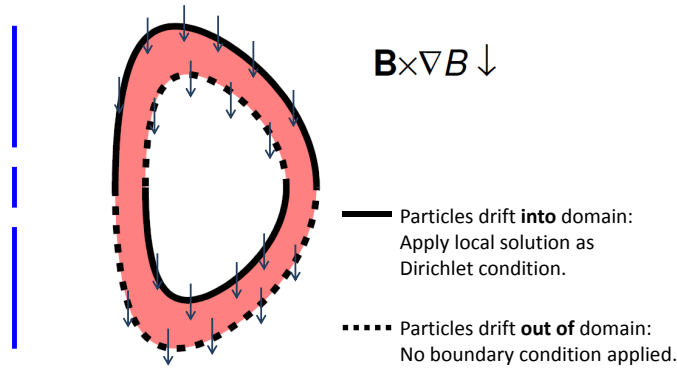}
\caption{The simulation domain is an annulus in the poloidal plane (shaded).
A boundary condition in $\psi$ must be supplied wherever particle trajectories enter the domain
(thick solid curves).  Where trajectories exit the domain (thick dashed curves),
one-sided differentiation is used so no boundary condition
is imposed.
\label{fig:boundaryConditions}}
\end{figure}

As the first derivative $\partial /\partial \psi$ appears in the kinetic equation, we must
supply boundary conditions in the $\psi$ coordinate where the
collisionless trajectories enter the domain.
Supposing the $\nabla B$ and curvature drifts are downward, this means a boundary condition
should be supplied on the top half of the outer $\psi$ boundary, and on the bottom half of the
inner $\psi$ boundary (Fig. \ref{fig:boundaryConditions}).
If the drifts are upward rather than downward, the boundary conditions are reversed appropriately.

To obtain these boundary conditions, the local $\delta f$ equation (\ref{eq:local}) is first solved
on the entire inner and outer boundaries.
The resulting distribution functions are then imposed as Dirichlet boundary
conditions for the global solution,
but only on the half-boundaries where trajectories enter the domain.
On the half-boundaries where trajectories exit the domain, the global drift-kinetic equation
is imposed using a one-sided (i.e. upwinded) $\partial/\partial \psi$ derivative.

Even with these boundary conditions, the
solutions of the kinetic equation with no sources ($S_a=0$) are not
well behaved. The flux surface averaged
density and pressure carried by $g_a$ will
generally be nonzero,
and the approximation $|g_a| \ll |\fMa|$ can be violated.
Physically, the radial particle and heat fluxes which arise from
the specified equilibrium profiles in the $\delta f$ approximation
will not generally be consistent with those equilibrium profiles, in the absence of sources/sinks.
When $S_a=0$, the code will find an unphysical large $g_a$ to force
the radial fluxes to be consistent with density and pressure profiles of
the total $f_a$.
This behavior may be understood in light of Section 8 of Ref. \cite{usPedestal},
in which it is proved that generally no time-independent solution exists of the global $\delta f$
kinetic equation when $S_a=0$.
For similar reasons, sources/sinks are also mandatory in global $\delta f$ turbulence codes, and a variety
of forms have been used
\cite{GYROSources, GENESource, globalGENE}.
A closely related issue was also discussed in Ref. \cite{PerPotato} in the context of transport
near the magnetic axis.

For the neoclassical problem considered here, the time-independent kinetic equation becomes solvable
(and has sensible solutions) when posed in the following manner.
We allow $S_a$ to include both particle and heat sources
of magnitudes that are initially unknown.  The kinetic equation is then solved,
subject to the additional constraints that the flux surface averaged density
and pressure be contained purely in $\fMa$ and not in $g_a$:
\begin{eqnarray}
\left< \int d^3v\; g_a \right>&=&0,
\label{eq:gNoDensity}
\\
\left< \int d^3v\; v^2 g_a \right> &=&0.
\label{eq:gNoPressure}
\end{eqnarray}
The additional constraints (\ref{eq:gNoDensity}) - (\ref{eq:gNoPressure})
allow us to solve for the unknown particle and heat source profiles.
These profiles can be thought to represent additional transport (e.g. the divergence of turbulent fluxes)
such that, when added to the neoclassical transport associated with $g_a$,
the total transport is consistent with the chosen equilibrium profiles,
i.e. the total fluxes are independent of radius.

To implement this method, we choose
\begin{equation}
\label{eq:source}
\hat{S}_a = S_{pa}(\psi) \Theta(\theta) \left( x_a^2-\frac{5}{2}\right)e^{-x_a^2}
+S_{ha}(\psi) \Theta(\theta)\left( x_a^2 - \frac{3}{2}\right) e^{-x_a^2},
\end{equation}
where $x_a = v/ v_a$ is the speed normalized to the thermal speed $v_a=\sqrt{2T_a(\psi)/m_a}$,
and $\hat{S}_a$ is the source normalized as described in the appendix.
The function $\Theta(\theta)$ is
typically chosen to be either $\Theta=1$ or $\Theta = 1 + \cos\theta$, the latter motivated by the ``ballooning" (outboard-localized)
character of turbulent transport.
The form (\ref{eq:source}) is chosen
so that when $\int d^3v$ and $\int d^3v\;v^2$ are applied,
the $S_{pa}$ term gives rise to a particle source but not a heat source,
and the $S_{ha}$ term gives rise to a heat source but not a particle source.
If there are $N_\psi$ radial grid points,
the profiles $S_{pa}(\psi)$ and $S_{ha}(\psi)$
represent a total of $2N_\psi$ additional unknowns per species, and the constraints
(\ref{eq:gNoDensity}) - (\ref{eq:gNoPressure}) at each flux surface represent a total of $2N_\psi$ additional conditions per species,
so the overall linear system remains square.
For the case of a single species of ions, the linear system to solve becomes
\begin{equation}
\begin{array}{r}
\mbox{Kinetic equation}\;\;\{ \\
\left<\int d^3v g_i\right>=0 \;\;\{ \\
\left<\int d^3v\; g_i v^2\right>=0\;\; \{
\end{array}
\left(\begin{array}{ccc}
M_{11} & M_{12} & M_{13} \\
M_{21} & 0 & 0 \\
M_{31} & 0 & 0 \end{array}
\right)
\underbrace{
\left(
\begin{array}{c}
g_i \\
S_{pi} \\
S_{hi} \end{array}
\right)}_{\mbox{Vector of unknowns}}
 = \left(
\begin{array}{c}
-\vect{v}_{\mathrm{m}i}\cdot\nabla f_{\mathrm{m}i}
 \\
0 \\
0 \end{array}
\right),
\label{eq:blocks}
\end{equation}
where $-\vect{v}_{\mathrm{m}i}\cdot\nabla f_{\mathrm{m}i}$ is the right-hand side from (\ref{eq15}),
and the $M$ operators are as follows: $M_{11} = (\vpar\vect{b} + \vdao)\cdot\nabla - C_{\ell a}\{ \ldots \}$ is the source-free drift-kinetic operator,
$M_{12}$ and $M_{13}$ represent the $S_{pi}$ and $S_{hi}$ terms
in (\ref{eq:source}) respectively,
and $M_{21}$ and $M_{31}$ represent (\ref{eq:gNoDensity}) - (\ref{eq:gNoPressure}) respectively.
For the case of multiple particle species, the linear system consists of blocks of the form (\ref{eq:blocks}) for each
species, with coupling between species only through the collision operators in the $M_{11}$ blocks.

Sources/sinks are not required in a time-dependent (initial-value) simulation. However, in a time-dependent global code
without sources, the user must make an arbitrary choice of when to terminate the simulation,
since a true steady state does not exist. Therefore the approach used here is no more arbitrary.
The time-independent approach is also faster, especially at low collisionality, since
convergence is not limited by the rate of physical dissipation.

Since the form of the sources (\ref{eq:source}) is not rigorously derived, it is useful
to consider the conditions under which the sources are minimized.
The particle source tends to be smaller than the heat source, since
for a single ion species in the local limit, it can be shown that the particle flux vanishes.
In the global case for one species, or for multiple species, the particle flux no longer needs to be exactly zero,
meaning some particle source is needed, but in practice
it often remains small compared to the heat source.
The heat source can be minimized if the density and temperature input profiles
give rise to a heat flux which is nearly independent of $\psi$.
In the banana and \PS collisionality regimes,
the heat flux scales as $\sim n_i^2 dT_i/d\psi$,
and so the heat source is minimized when $n_i^2 dT_i/d\psi \sim$ constant.
In the plateau regime,
the heat flux scales as $\sim n_i\; dT_i/d\psi$,
and so the heat source is minimized when $n_i\; dT_i/d\psi \sim$ constant.

The linear system is discretized using independent variables  $(\psi_N, \theta, x_a, \xi)$
where $\psi_N = \psi / \psi_a$ is the normalized flux, $\psi_a$ is the flux at the last closed flux surface,
and $\xi = \vpar/v$ is the cosine of the pitch angle.
Note that the speed normalization $v_a$ depends on $\psi$ through the temperature.
In the $\psi_N$ coordinate, we employ a grid
with finite difference derivatives using a 5-point stencil.
In $\theta$, we employ a grid with either
a 5-point finite-difference stencil
or a spectral differentiation matrix \cite{Trefethen}.
In $x_a$, we use the spectral collocation discretization
described in Ref. \cite{speedGrids}.
This discretization employs a nonuniform grid for the distribution function, together
with a uniform grid for the Rosenbluth potentials in the collision operator.
The collision operator and Rosenbluth potentials are implemented in the code
as detailed in Ref. \cite{speedGrids}.
In the $\xi$ coordinate, we use a modal approach, expanding in Legendre polynomials $P_L(\xi)$.
The appendix details the form of the kinetic equation when these independent variables
are used.
Up-down asymmetry in the magnetic geometry is allowed, so no symmetries are assumed in the distribution
function \cite{WongChan, usPedestal}.

A useful test of the discretized operators in PERFECT is the following. If the collisionality
is set to zero, the remaining terms in $M_{11}$ should conserve $W_{a0}$, $\mu$, and the
gyroaveraged canonical momentum $\psi_{*a} = \psi-I\vpar/\Omega_a$.
Thus, the vectors corresponding to
$W_{a0}$, $\mu$, and $\psi_{*a}$, when multiplied by
the matrix $M_{11}$, should give zero to within discretization error.  It was verified that the code passed this
test even when the input fields and profiles ($B(\theta,\psi)$, $\Phi_0(\psi)$, etc.) were complicated functions.

To solve the large sparse linear system (\ref{eq:blocks}), we use a Krylov-space
iterative method. Preconditioning is essential, or else the Krylov solver will not converge in a
reasonable number of iterations.
To obtain an effective preconditioner, the matrix $M_{11}$ in (\ref{eq:blocks}) may be simplified in several ways.
(We find that simplifying $M_{21}$, $M_{31}$, $M_{12}$, or $M_{13}$ in the preconditioner does not lead to convergence.)
One effective option is to drop all elements of $M_{11}$ that are off-diagonal in the $x_a$ coordinate.
Another option is to use a 3-point stencil instead of the 5-point stencil for radial and/or poloidal derivatives.
Also, the terms that are pentadiagonal in the Legendre polynomial $P_L(\xi)$ index $L$ may be dropped, leaving the preconditioner tridiagonal in $L$.
Depending on the physical and numerical parameters, some of these preconditioning options may not lead to convergence.
In practice, we find the most robust preconditioner is obtained by dropping terms that are off-diagonal
in $x_a$ but retaining the full coupling in the other three coordinates.
The preconditioner is $LU$-factorized directly using either MUMPS\cite{MUMPS:1} or SuperLU-dist\cite{superlu1, superlu2}.
Typically, either GMRES\cite{GMRES} or BICGStab(l)\cite{BICGstabl} is then effective at iterating
to a solution of the full system.
Independent versions of the code have been developed in Matlab and in Fortran/PETSc\cite{petsc-web-page, petsc-user-ref}.
It was verified that the two codes produce the same output when given identical inputs.

When the $\vdao\cdot(\nabla g_a)_{\mu,W_{a0}}$ term is turned off, it can be shown that $S_{pa}=0$ and $S_{ha}=0$, and
the kinetic equation solved by PERFECT reduces to the local $\delta f$ equation solved by conventional neoclassical codes.
We have therefore benchmarked PERFECT in this limit against the local codes from Ref. \cite{WongChan} (one species only) and Ref. \cite{NEOFP}
(considering cases of one, two, and three species) for a range of collisionalities. In all cases examined, the results of these codes for the flows and radial fluxes agreed precisely.

\section{Summary of analytic results}
\label{sec:plateau}

In this section we discuss the analytic formulae to which we compare
the numerical results. For this section and the next, we drop the species subscript to simplify notation
since all quantities refer to the single ion species.

Analytic expressions for the ion flow and heat flux in a plateau-regime pedestal,
including finite orbit with effects, were derived in Refs. \cite{Istvan, Peter}.
These calculations exploit expansions in collisionality and aspect ratio,
and assume circular concentric flux surfaces.
Conservation of $\psi_*$ is used to retain variation in $\Phi_0(\psi)$ and $v$
in the kinetic equation as a particle moves in $\theta$.
A Krook approximation for the collision operator is made after
subtracting the mean flow from the distribution function.

In both references \cite{Istvan, Peter}, the ion heat flux is found to be
\begin{eqnarray}
\label{eq:plateauQ}
\left< \vect{q}\cdot\nabla\psi\right>
&=& \left< \int d^3v \frac{mv^2}{2} g \vm\cdot\nabla\psi\right> \\
&=& \left<\vect{q}\cdot\nabla\psi\right>_{local}
\frac{1}{3}
\left(
\frac{
4U^8 + 16 U^6 + 24 U^4 + 12 U^2 + 3}
{2U^4 + 2U^2 + 1}
\right)
e^{-U^2} \nonumber
\end{eqnarray}
where
\begin{equation}
\left<\vect{q}\cdot\nabla\psi\right>_{local}
= - \frac{3\sqrt{\pi} \epsilon^2 I^2 n v_i^3}
{4q R \Omega^2}
\frac{dT}{d\psi}
\end{equation}
is the plateau-regime heat flux in conventional local theory,
$u = cI B^{-1}\; d\Phi_0/d\psi$, and $U = u/v_i$
is the poloidal Mach number.
Also, in equations (64)-(65) of Ref. \cite{Peter},
the parallel flow is found to be determined by a radial differential equation:
\begin{equation}
\frac{1}{p}\frac{dp}{d\psi} + \frac{e}{T}\frac{d\Phi_0}{d\psi}
+\frac{eB}{TcI}V_{||}
-u \frac{d}{d\psi} \left( \frac{m V_{||}}{T}\right)
+\left( \frac{4U^6-2U^4+1}{2U^4+2U^2+1}\right) \frac{1}{2T}\frac{dT}{d\psi}
=0.
\label{eq:flow}
\end{equation}
(In ref \cite{Peter} this equation is written in terms of $k$ rather than $V_{||}$,
but the two versions are related by (\ref{eq17}), noting $r_{T} \gg \rhop$ and $r_{\eta} \gg \rhop$ and retaining
$dk/d\psi$.)
The poloidal variation of $V_{||}$ and $k$ are $O(\epsilon)$ so both quantities may be treated
as flux functions in (\ref{eq:flow}).

In the local limit,  $U\ll 1$, so $\left< \vect{q}\cdot\nabla\psi\right> \to \left<\vect{q}\cdot\nabla\psi\right>_{local}$. Also,
in the local limit the $d(mV_{||}/T)/d\psi$ term in (\ref{eq:flow}) becomes
formally small, so (\ref{eq:flow}) reduces to (\ref{eq17}) with $k=-1/2$,
recovering the conventional local result.

There are several shortcomings of the analytic calculations in
Refs. \cite{Istvan, Peter} leading to (\ref{eq:plateauQ}) and (\ref{eq:flow}). Primarily, it is not strictly
allowed within these analytic calculations to have a heat flux which is independent of $\psi$,
since the heat flux scales as $\sim nT^{3/2}\,dT/d\psi$, while $r_n \sim \rho_\theta$ and $\rTi \gg \rhop$.
Thus, either a source/sink should be included in the analytic calculation,
or else $d\Ti/d\psi$ must be allowed to
vary on the $\rhop$ scale (i.e. $d^2\Ti/d\psi^2$ is large) to allow the heat flux to be
$\psi$-independent.  (In this latter scenario,
$dT/d\psi$ is much larger at the bottom of the density pedestal than at the top, so
the product $nT^{3/2}\,dT/d\psi$ remains constant.)
These features have now been incorporated into the analytic theory,
details of which will be presented in a separate publication.
The updated analytic theory also allows for substantial ion particle flux (on the ion gyro-Bohm level),
a possibility which was not included in earlier analytic work.  This possibility may be important
if different radial boundary conditions are used in the code to allow large particle fluxes through the domain,
but we do not consider this possibility further here.
When the ion neoclassical particle flux is thus assumed to be negligible, the results (\ref{eq:plateauQ}) and (\ref{eq:flow})
turn out to be unchanged, even in the presence of large $d^2\Ti/d\psi^2$.
We therefore expect agreement between (\ref{eq:plateauQ})-(\ref{eq:flow})
and PERFECT numerical calculations in the appropriate regime of aspect ratio, collisionality, and
circular flux surface shape.

\section{Comparison of numerical and analytic calculations}
\label{sec:comparison}

We now describe the paradigm used for comparing the numerical and analytic solutions.
We continue to suppress species subscripts since only a single ion species is considered
in the analytic calculation.

While the code allows for a general shaped magnetic equilibrium with closed flux surfaces,
for comparison with analytic theory
here we use the following simplified magnetic geometry: $B(\psi,\theta) = \bar{B}/(1+\epsilon\cos(\theta))$,
$\vect{b}\cdot\nabla\theta = 1/(q \bar{R})$, $q=3$, and $I = \bar{B}\bar{R}$,
with $\bar{R}$ and $\bar{B}$ the normalization constants described in the appendix.

In principle, any profiles of density, temperature, and potential may be specified,
as long as the resulting distribution function does not violate the $|g| \ll \fM$ ordering.
For comparison to the analytic theory, profiles are specified as follows.
First, analytic functions are chosen for $\Phi_0(\psi)$ and $\eta(\psi)$.
For results shown here,
$\eta=$constant unless stated otherwise. Then the remaining profiles $n(\psi)$
and $T(\psi)$ are found by numerical solution of the following pair of coupled nonlinear 1D equations:
$n=\eta\exp(-e\Phi_0/T)$ and $Q=$constant where $Q(n,T,d\Phi_0/d\psi)$ is the
analytic heat flux (\ref{eq:plateauQ}).
This latter equation is imposed to minimize any possible influence of the source term.
The agreement between the analytic and numerical calculations
is robust and persists even when other choices are made for the input profiles,
but we focus on the constant-$Q$ case here since the agreement is then particularly precise.

The inverse aspect ratio $\epsilon$ is chosen to be $0.001$ so the $\epsilon \ll 1$ approximation
made in the analytic theory should be well satisfied, and so a large plateau
regime should exist between the banana and \PS regimes.
Unless otherwise specified, results were obtained using poloidally symmetric sources, $\Theta(\theta)=1$ in (\ref{eq:source}).

\begin{figure}[h!]
% Figure generated using m20130604_01_figureOfPlateauProfilesAndConvergence.m
% Data: PERFECTOutput_20130604-03_convergence_merged.h5
% Merged using m20130119_01_mergeMorpheusHDF5Files.m
% from PERFECTOutput_20130604-01_convergence_easy.h5
% and PERFECTOutput_20130604-02_convergence_moreNx.h5.
% Runs on Hopper in
% ~/20130529-02-perfect/convergenceFigure/
%*****************************
% Original figure generated using:
% m20130203_02_figureOfPlateauProfilesAndConvergence.m
% Data generated using:
% loki:20130126-01-morpheus/convergenceFigure
% Data:
%
\includegraphics{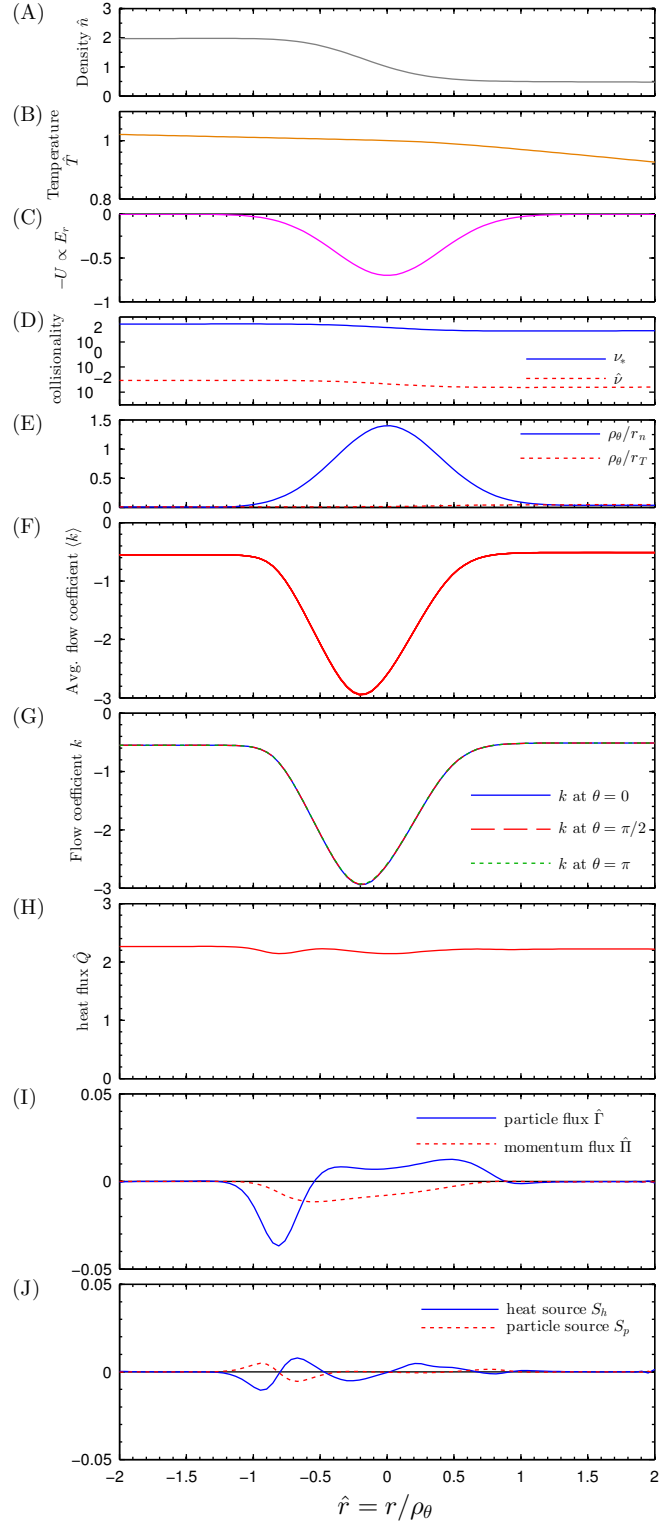}
\caption{(Color online) Input and output profiles for the plateau regime numerical
computation at $\epsilon=0.001$, for a target value of $U=0.7$ in the middle of the pedestal.
\label{fig:plateauProfilesAndConvergence}}
\end{figure}

Figure \ref{fig:plateauProfilesAndConvergence}.A-C shows the input profiles used.
The horizontal coordinate
$\hat{r}$ is defined by
\begin{equation}
\frac{d\hat{r}}{d\psi} = \frac{e}{m c I v_i}\left< \hat{B}^2 \right> ^{1/2}
\end{equation}
where $\hat{B} = B/\bar{B}$,
so that $\hat{r} \sim r / \rhop$ where
$r$ is the minor radius.  The point $\hat{r}=0$ is an arbitrary
location in the middle of the pedestal.
The potential specified is $\hat{\Phi}(\psi_N) = \Phi'
\erf(s[\psi_N-\psi_{N0}])$
where $\psi_N$ is the normalized flux,
and $\psi_{N0}$, $s$,
and $\Phi'$ are constants.
The $n$ and $T$ profiles are computed as described already so that the anticipated heat flux (\ref{eq:plateauQ}) is independent of $\psi$.
We are free to choose an integration constant in selecting $T(\psi)$ by the procedure described earlier,
and for results here we choose $(1/T) dT/d\psi_N = -0.2$ in the middle of the domain.
This choice gives a very gentle $T$ profile compared to the $n$ profile,
so the orderings used in the analysis will be well satisfied.
Indeed, Figure \ref{fig:plateauProfilesAndConvergence}.E shows
$\rhop/\rT \ll 1$, while $\rhop / \rn$ is $O(1)$ in the middle of the pedestal.
As shown in Figure \ref{fig:plateauProfilesAndConvergence}.D, the collisionality is chosen
so that the plateau approximation is well satisfied throughout the domain:
$\nuPrime \ll 1 \ll \nu_*$.

Figure \ref{fig:plateauProfilesAndConvergence}.F
shows the flux surface averaged parallel flow coefficient computed by the code.
It is crucial to verify that outputs such as this quantity are well converged with respect
to the many numerical parameters.
These parameters include
the number of radial grid points $N_\psi$,
the number of poloidal grid points $N_\theta$,
the number of Legendre modes retained for the distribution function $N_\xi$,
the number of Legendre modes retained for the Rosenbluth potentials $N_p$,
the number of speed grid points for the distribution function $N_x$,
the number of speed grid points for the Rosenbluth potentials $N_y$,
the maximum $x$ value for the Rosenbluth potentials $x_{\mathrm{max}}$,
the width of the domain in $\psi$,
and the tolerance used to define convergence of the Krylov solver.
To verify convergence, in Fig. \ref{fig:plateauProfilesAndConvergence}.F
we plot the results of 10 runs, consisting of a base case and 9 runs in which each resolution parameter
is doubled in turn (or in the case of the solver tolerance, reduced by $10\times$.)
The results are indistinguishable, demonstrating the excellent convergence.
The base case shown in Fig. \ref{fig:plateauProfilesAndConvergence}
was computed with the following resolution parameters:
$N_\psi = 61, N_\theta=5, N_\xi=65, N_p = 4, N_x=12, x_{\mathrm{max}}=7,$ and $N_y=350$.
Computation for the base case took two minutes on a single node of Hopper at NERSC (containing 24 cores).

As noted previously, in the pedestal, the parallel flow coefficient $k$ can vary poloidally,
in contrast to conventional local theory.
However, based on the analytic theories, the poloidal variation of $k$ is expected to become small as $\epsilon$ decreases \cite{Peter}.
Figure \ref{fig:plateauProfilesAndConvergence}.G
shows $k$ at the outboard $(\theta=0)$ and inboard $(\theta=\pi)$ midplanes and at
the top of the flux surface $(\theta=\pi/2)$.
The poloidal variation of $k$ is indeed very small due to the small value of $\epsilon$.

Figure \ref{fig:plateauProfilesAndConvergence}.H shows the radial heat flux,
normalized as detailed in the appendix.
The heat flux is nearly constant
since the profiles were constructed to have constant heat flux
according to the analytic theory. The numerical heat flux does have some very small
radial variation so it evidently differs slightly
from this predicted value, though not by a large relative amount.
Figure \ref{fig:plateauProfilesAndConvergence}.I shows the particle and momentum
fluxes computed by the global code.  These fluxes are both exactly zero
in a local calculation.  In a global calculation they need not be zero,
though they tend to be quite small compared to the heat flux,
as can be seen by comparing the magnitudes of plots (I) and (H).
The particle and heat sources needed to maintain the profiles
in steady state are given in Figure \ref{fig:plateauProfilesAndConvergence}.J.
The normalizations used for the fluxes and sources are given in the appendix.

In figure \ref{fig:PetersProfiles},
the flow and heat flux are compared for several models, considering
the same input profiles, repeated in figure \ref{fig:PetersProfiles}.A-C for convenience.
Figure \ref{fig:PetersProfiles}.D compares profiles of the $dT/d\psi$-driven part of the flow, $k$.
The global numerical result (solid red curve) shows a depression in $k$,
in contrast to the local analytic prediction $k=-1/2$.
A local numerical calculation is plotted,
using PERFECT to retain finite $\epsilon$ and finite $\nu$,
but turning off the $\vdo\cdot(\nabla g)_{\mu,W_0}$ term in the code.
This local calculation shows a slight departure from $k=-1/2$ due to the small change in collisionality
across the pedestal, but the variation in $k$ in this local calculation
is negligible compared to the variation
in the global calculation.
Also plotted is the solution of the 1D equation
(\ref{eq:flow})
from the global analytic calculation.
The  global numerical and global analytic calculations agree very closely.
Lastly, plotted in blue is a PERFECT calculation in which the poloidal drift (dominantly $\ExB$) is retained but the radial
derivative of $g$ is dropped. It might be hoped that this approach would capture much of the physics, while
eliminating the radial coupling that makes the numerical solution challenging. However,
the blue curve is quite different from the full global calculation (red), so
it is evidently not a good approximation to drop $\partial g/\partial \psi$ even when $\epsilon \ll 1$.

The heat fluxes for the same four models are plotted in figure \ref{fig:PetersProfiles}.E.
Due to the way the input profiles are generated, the global analytic heat flux in figure \ref{fig:PetersProfiles}.E
is necessarily constant.
The heat flux from the global numerical calculation
is nearly constant, but with a slight deviation.
The heat flux calculated in the local kinetic model is lower.
Again, neglecting the $\partial g/\partial \psi$ term in the global calculation
leads to a very different and unphysical result.

\begin{figure}[h!]
% Figure generated using m20130528_02_figureOfPetersProfilesInPlateauForPaper.m
% Code output files used: (in my MATLAB directory)
% PERFECTOutput_20130529-49_PetersProfiles2_local.h5
% PERFECTOutput_20130529-50_PetersProfiles2_global.h5
% Code run on Hopper:
% ~/20130529-02-perfect/PetersProfilesFigure/2/local
% ~/20130529-02-perfect/PetersProfilesFigure/2/global
% *************************************
% Previous version: Figure generated using
% m20130201_01_figureOfPetersProfilesInPlateauForPaper.m
% Data generated using:
% loki:20130201-01-morpheus/PetersProfilesNewNu
% Data:
% morpheusOutput20130202_02_PetersProfilesNewNu_local.h5
% morpheusOutput20130202_03_PetersProfilesNewNu_baseCase.h5
% morpheusOutput20130202_01_PetersProfilesNewNu_woddpsi.h5
% Convergence tests for these parameters:
% loki:20130201-01-morpheus/PetersProfiles/
%\includegraphics{PetersProfilesInPlateauForPaper.eps}
\includegraphics{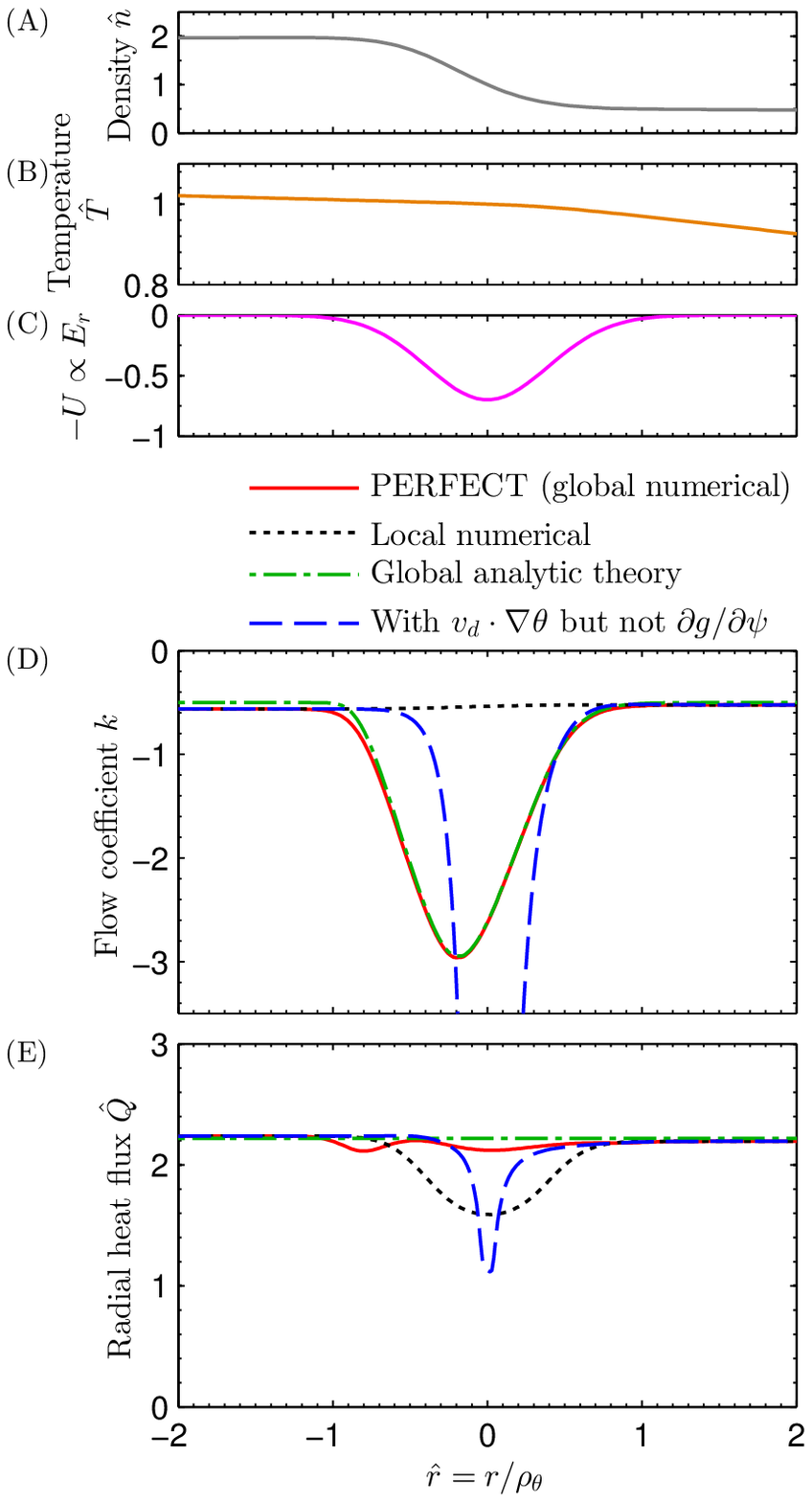}
\caption{(Color online) The flow and heat flux computed by the new global code
PERFECT closely match the predictions of the analytic theory in Section \ref{sec:plateau}.
Input profiles are shown in (A)-(C).
In (D), the ``analytic theory'' curve is obtained by solution of
(\ref{eq:flow}).
A depression in $k$ is both predicted by the analytic theory and observed in the global
numerical calculation, in contrast to the constant result $k=-1/2$ from local theory.
A local kinetic calculation shows slight variation from $k=-1/2$ due to variation
in collisionality across the pedestal.
The input profiles are chosen to have constant heat flux according to the global
analytic theory (\ref{eq:plateauQ}), shown in (E), and a nearly constant heat flux is indeed observed in the global numerical solution.
In contrast, the local calculation predicts a lower heat flux in the pedestal.
The blue dashed curve is computed by retaining the poloidal drift but neglecting
the $\partial g/\partial \psi$ term, and it is evident from (D)-(E) that this
approach does not yield a good approximation of the correct (red) result.
\label{fig:PetersProfiles}}
\end{figure}

\begin{figure}[h!]
\includegraphics{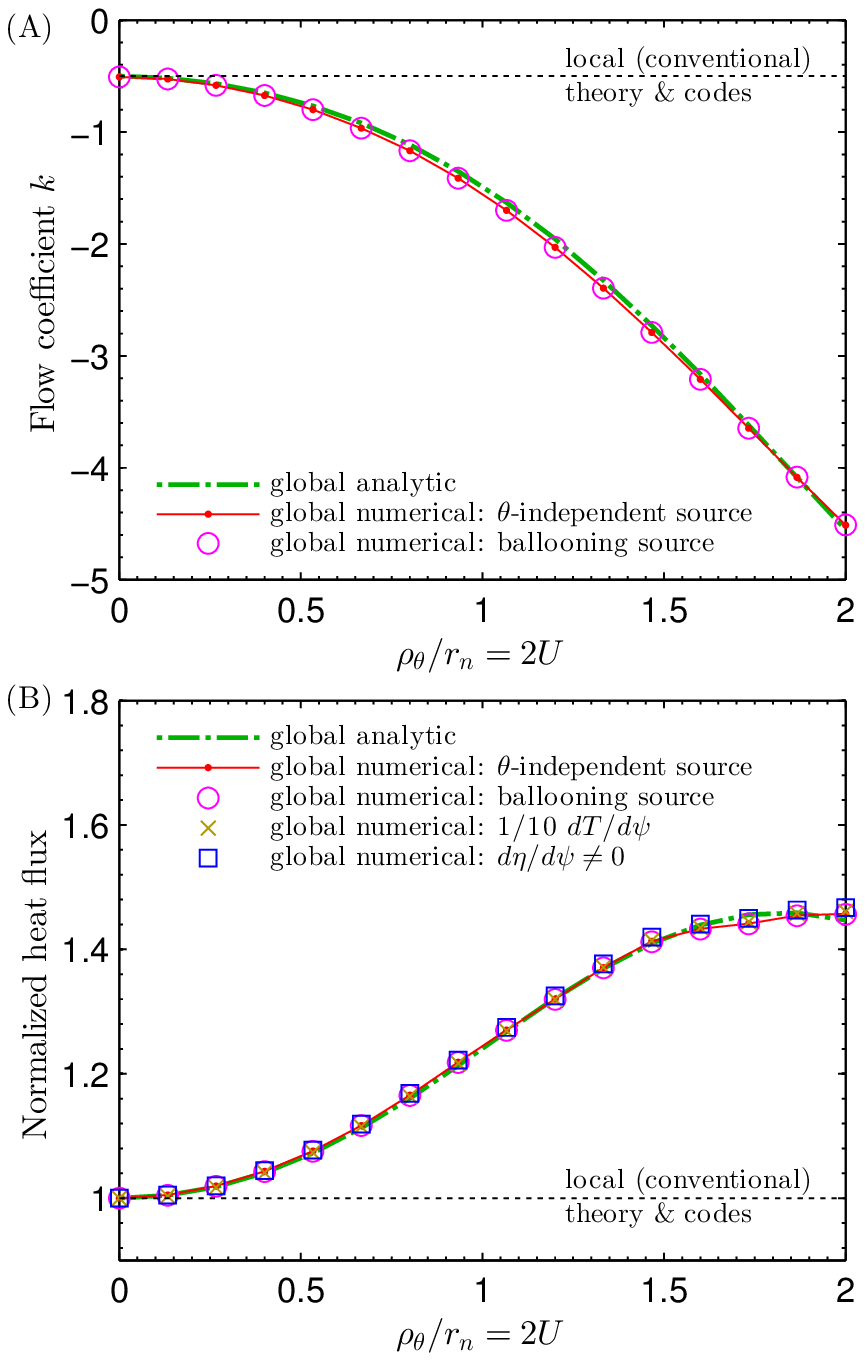}
\caption{(Color online) The new global code and the analytic theory of section \ref{sec:plateau} agree precisely
for both the flow and heat flux
in the pedestal, with significant departures from the local values.
Each point in the graphs corresponds to a different set of input density, temperature, and potential profiles,
with the horizontal axis corresponding to the mid-pedestal value of the normalized density gradient. In (A), the vertical axis is the flow coefficient
in the middle of the pedestal. In (B), the vertical axis is the mid-pedestal heat flux normalized by the local value of the heat flux
at that location.
The numerical results are insensitive to the form of the source $\Theta(\theta)$.
In (B), we also show the numerical heat flux when normalized in this manner is
insensitive to the choice of $dT/d\psi$ or to $\eta(\psi)$,
as predicted by the analytic theory.
\label{fig:plateauComparison}}
\end{figure}

Using a series of numerical calculations resembling that in figure \ref{fig:PetersProfiles}
but for input profiles with different gradients,
the comparison in
Figure \ref{fig:plateauComparison} is generated.
The horizontal coordinate is the normalized density gradient at the middle
of the pedestal, defined by the extremum of the Gaussian-shaped $d\Phi_0/d\psi$ profile.
The flow coefficient from the same location
is plotted in Figure \ref{fig:plateauComparison}.A
for the global numerical, global analytic, and local models.
The agreement between the global numerical and analytic results is remarkable,
and both show a substantial departure from the local result.
We also repeat the scan
use a ballooning source $\Theta(\theta) = 1+\cos\theta$ in the global kinetic calculation in place of $\Theta=1$.
This change has negligible effect on the numerical results, since $\epsilon \ll 1$.
For larger $\epsilon$, the choice of poloidal variation for the source will generally have some effect
on the results.
In figure \ref{fig:plateauComparison}.B, the mid-pedestal heat fluxes
are plotted, normalized by the local formula.
Again the agreement between the global analytic and numerical calculation is excellent, and the alternative
form of the source yields results that are nearly indistinguishable.
We also vary two other aspects of the profiles to demonstrate the insensitivity of the global numerical results.
First, the mid-pedestal value of $(1/T) dT/d\psi_N$ (which was a free parameter when determining the input profiles) is reduced by a factor of 10.
Second, the density profile is altered by choosing $\eta(\psi_N) = 1+0.1(\psi_N - \psi_{N0})$ instead of $\eta = 1$,
introducing a departure from electrostatic ion confinement.
The analytic theory predicts the normalized heat flux should not be altered by these changes, and this
prediction is borne out by the global numerical calculations.

\section{More realistic conditions}
\label{sec:advanced}

\begin{figure}[h!]
% Figure generated using m20130605_03_figureOfPlateauProfilesAndConvergence_bigEpsilon.m
% Data files:
% PERFECTOutput_20130605-10_bigEpsilonConvergence.h5
% Code run on Loki
% ~/20130529-01-perfect/bigEpsilon/8/
\includegraphics{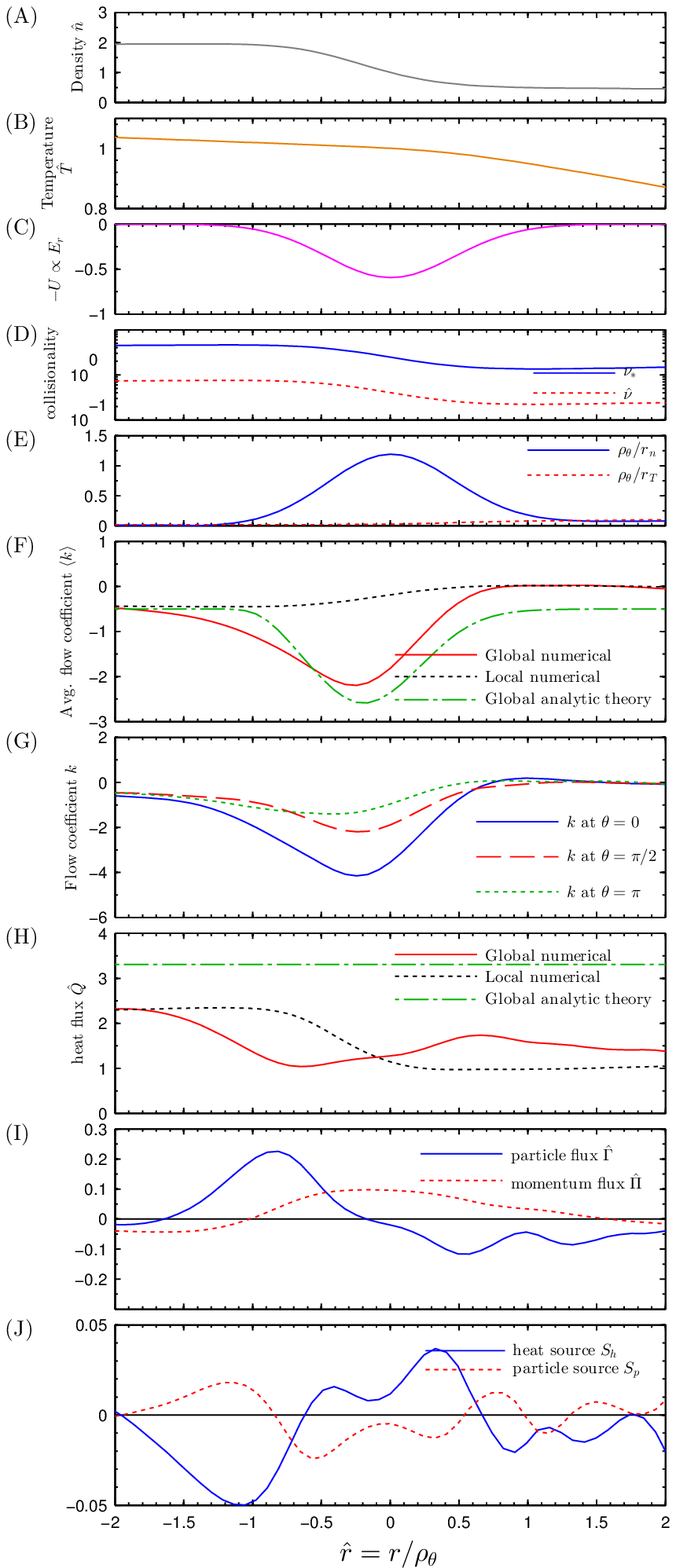}
\caption{(Color online) Same as Figure \ref{fig:plateauProfilesAndConvergence}, but for D-shaped flux surfaces with $\epsilon=0.3$.
In (F) and (H), the global analytic theory no longer precisely matches the global numerical calculation
as it did in figure \ref{fig:PetersProfiles},
although the theory correctly predicts a depression in $k$.
Notice in (G) that the poloidal variation of $k$ is now significant.
\label{fig:bigEpsilon}}
\end{figure}

To examine the changes to the results when more realistic magnetic geometry is used,
figure \ref{fig:bigEpsilon} shows inputs and outputs of PERFECT when Miller geometry\cite{Miller}
is used, with $\epsilon=0.3$. (Other Miller parameters are as given in Ref. \cite{Miller}:
$\kappa=1.66, \delta = 0.416, s_\kappa = 0.70, s_\delta = 1.37, \partial R_0/\partial r = -0.354$,
and $q=3.03$).
The collisionality is chosen so the entire domain remains in the plateau regime, though
the plateau regime is not well defined since $\epsilon$ is so large.
A well is again observed in the flow coefficient $k$,
although the global numerical calculation no longer precisely agrees with the global analytic
calculation. This disagreement is not unexpected, given that the analytic theory is based on an expansion
in $\epsilon \ll 1$.
Notice in figure \ref{fig:bigEpsilon}.G that $k$ now has significant poloidal variation,
as discussed in Ref. \cite{usPedestal}.
In figure \ref{fig:bigEpsilon}.H, it can be seen that the analytic and numerical heat fluxes also
differ significantly at finite $\epsilon$.
The particle flux, momentum flux, and sources are significantly larger than in the $\epsilon =0.001$ case,
though the particle and momentum fluxes remain $< 10\%$ of the heat flux in normalized units,
and the normalized sources remain small.

\begin{figure}[h!]
% Figure generated using m20130920_06_figureOf3SpeciesPERFECTRunForPaper.m
% Data files:
% Runs from the matlab version, run both on my laptop and on edison in $GSCRATCH/20131120-PERFECT-matlab-Miller-3Species
\includegraphics{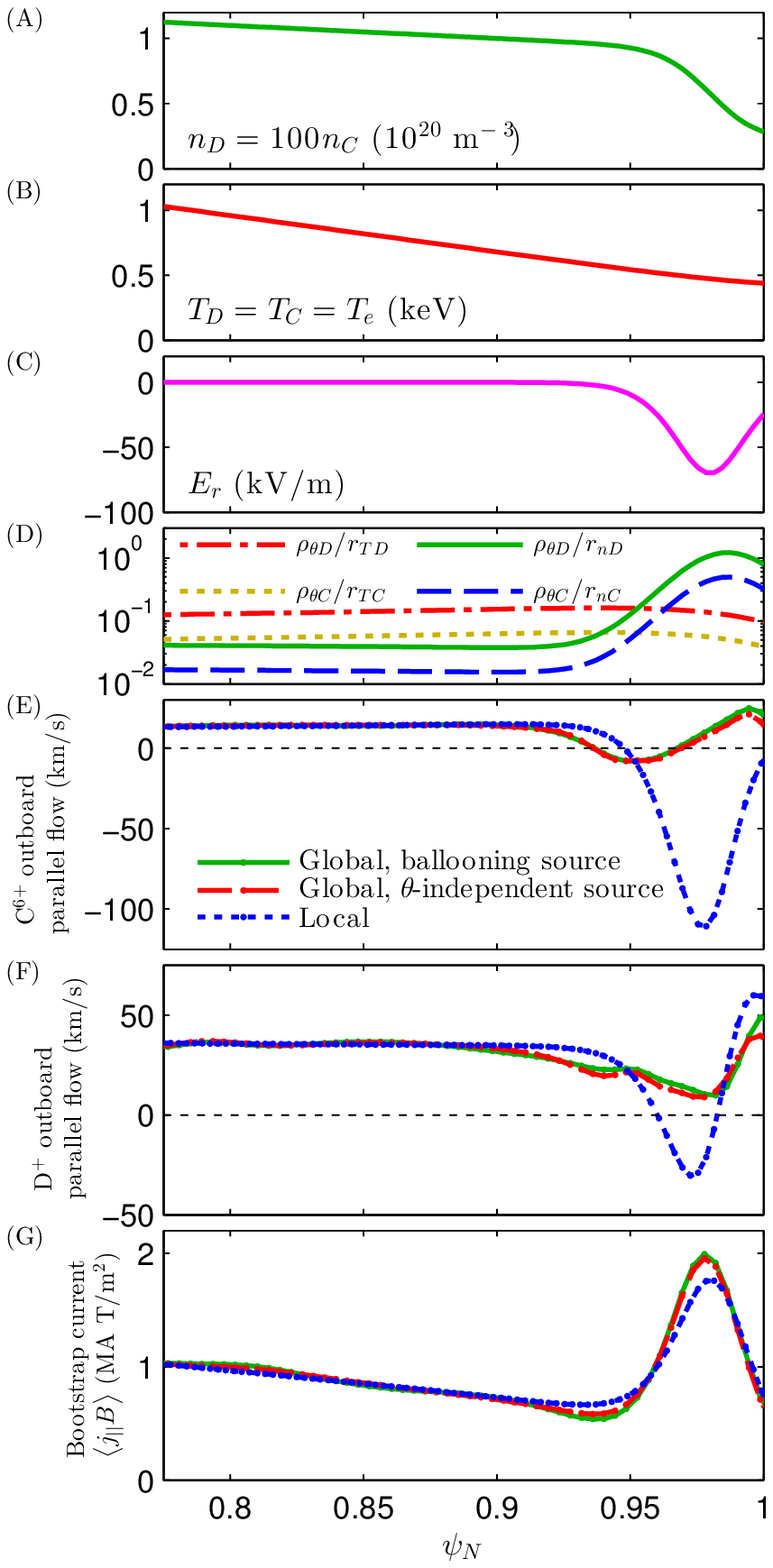}
\caption{(Color online) A simulation with three kinetic species. (A)-(C) illustrate the input profiles.
The ratios of poloidal gyroradii to density and temperature scale lengths for the two ion species are
shown in (D). Code outputs are shown in (E)-(G), including the ion parallel flows at the outboard midplane
and the bootstrap current. The local and global methods yield identical results in the core but predict
substantially different ion flows in the pedestal.
\label{fig:3species}}
\end{figure}

Finally, figure \ref{fig:3species} shows a scenario with multiple species.
In this computation, parameters reflective of DIII-D were used, but it is not meant to be a detailed
model of any particular device or shot.  Deuterium, carbon $6+$, and electrons are included,
retaining all collisional coupling through linearized Fokker-Planck collisions.
No expansions in mass ratios or impurity charge are made.
The input profiles
are shown in figure \ref{fig:3species}.(A)-(C). The equilibrium temperatures of the three
species are taken to be equal. The input carbon density profile is everywhere $1\%$ of the deuterium density profile. D-shaped geometry with aspect ratio 3 is again used.
As shown in figure \ref{fig:3species}.(D), the poloidal deuterium gyroradius is comparable to the density scale length in the steepest part of the density profile.
Figures \ref{fig:3species}.(E)-(F) show the ion parallel flows at the outboard midplane, for both local and global
simulations. In the pedestal region, there is a significant difference between the flows in the local and global models. Small differences propagate a distance $\sim \rho_{\theta i}$ towards the core, which
is not surprising since ion orbits convey information over this distance, but beyond this
distance there is essentially no difference between the local and global results.
Global results are shown both for the assumption of a ballooning source and for a source independent of $\theta$.
Since the inverse aspect ratio and $\rho_{\theta i}/r_{Ti}$ are only modestly smaller than 1, here the choice of source
has some visible effect on the results.  Without better understanding of the phase-space structure of the
turbulent fluxes that the source represents, it is unclear whether the ballooning or $\theta$-independent assumption is more accurate,
and the differences between the red and green curves gives a sense of the range of possible outputs.
Even accounting for this uncertainty, the trend towards reduced flow magnitude in the pedestal is robust.
It is apparent that the global effects considered here can have a significant impact on neoclassical
ion and impurity flows, and they may be important to consider in interpreting experimental measurements
of pedestal flows.

As shown in figure \ref{fig:3species}.(G), the bootstrap current is modified, though the modification is modest,
on the order of $\rho_{\theta i}/r_{Ti}$, as anticipated analytically in Refs. \cite{GrishaPRL,usPedestal}.
Profiles of the total current (including the Pfirsch-Schl\"uter component) look similar to the flux-surface-averaged
profiles shown.

The simulations shown in figure \ref{fig:3species} used
$N_\psi = 75, N_\theta=15, N_\xi=13, N_p = 4, N_x=6$.
The global calculations shown took 10 minutes on 4 cores of a
Dell Precision laptop with Intel i7-2860 2.50 GHz CPU and 16 GB memory.
For comparison, these figures represent a factor $\sim 600,000$ fewer CPU-hours
than the requirements quoted for similar global neoclassical simulations in Ref. \cite{XGCBootstrap}.

\section{Discussion and conclusions}
\label{sec:conclusions}

In this work, we have demonstrated
a new global $\delta f$ numerical framework for neoclassical
calculations in the pedestal.
The formulation is novel in several regards.
By including the source functions $S_p$ and $S_h$
as additional unknowns in the linear system,
as described in section \ref{sec:numerical},
it is possible to solve a time-independent problem (as in conventional local neoclassical codes) rather than a
time-dependent one (as in previous global codes), thereby greatly reducing the computational cost required.
Our code exploits modern Krylov-space solvers and the efficient velocity discretization of Ref. \cite{speedGrids}.
The global $\delta f$ approach
described in section \ref{sec:model},
while less general than the full-$f$ approach, reduces the kinetic equation
to a single linear system, decoupled from the quasineutrality condition.
At the same time, the global $\delta f$ method allows
finite orbit width effects to be retained that are not
captured in the conventional (local $\delta f$) approach.

We have also demonstrated precise agreement between a first-principles global analytic
theory and direct numerical simulation
for the flow and collisional heat flux in the pedestal.
In doing so, we have both verified the correctness of the analytic theory within its domain
of applicability, and also verified the new code we describe.
The benchmarking paradigm used here is important and novel
in that it depends centrally on the finite orbit width,
as we examine departures from zero-orbit-width (local) theory.
The benchmarking procedure may be useful to other kinetic simulations
of edge plasma \cite{Lin1,Lin2,Wang1,CSChang1,Wang2,CSChang2,Wang3,Kolesnikov,ORB5, Xu1,Xu2,XGCBootstrap,Dorf}.

The code and global analytic theory both demonstrate that the flow and collisional heat flux
generally suffer order-unity changes relative to local calculations when $\rn$ becomes as small as
$\rhop$, even if $r_{Ti}$ remains $\gg \rhop$.
In local theory, the flow depends only on plasma parameters
and their gradients on that surface.
However, in the global theory, a radial differential equation rather than algebraic equation
is obtained for the flow, meaning the flow depends on plasma parameters
on adjacent flux surfaces.
It is perhaps not surprising that the
flow acquires this nonlocal  character,
as flux surfaces are able to communicate
in our finite-orbit-width ordering in a manner that is
excluded in conventional theory.

\changed{
As in global $\delta f$ turbulence calculations, sources/sinks of particles and heat are an unavoidable element
of a global $\delta f$ neoclassical calculation, since the computed fluxes generally vary radially, implying that no
time-independent equilibrium is possible without sources.  The sources can be made small by choosing
the input profiles of density, temperature, and electric field so as to make the fluxes as uniform as possible,
as we have done for the comparison to analytic theory in section \ref{sec:comparison}. A source $S(\psi)$ may
be written as the derivative of a flux $\Gamma = \int S\, d\psi$, and the sources
computed in our method may be thought of as the derivatives of anomalous fluxes. In our approach where $S_p$ and $S_h$
are computed at the same time as the distribution function, the code effectively computes what the anomalous fluxes would need to be
-- up to $\psi$-independent constants -- such that the total (neoclassical + anomalous) fluxes becomes $\psi$-independent,
allowing a steady state. As these constant offsets in the anomalous fluxes cannot be determined by our method, it is still
possible in our approach for the anomalous fluxes to be larger than the neoclassical fluxes.
(For example, a large $\psi$-independent anomalous flux gives no effective source.)
At the same time, it is necessary in our approach to assume some specific form for the $\theta$- and velocity-space dependence
of the anomalous fluxes/sources, and the form we have assumed may not accurately reflect the true phase-space structure of
the plasma turbulence.  One approach to addressing this limitation without directly simulating the turbulence
is to compare several options for the source, as
we have done in figures \ref{fig:plateauComparison} and \ref{fig:3species}.
}

One noteworthy feature of the global $\delta f$ model that is visible in figure \ref{fig:plateauProfilesAndConvergence}.I and \ref{fig:bigEpsilon}.I
is the presence of a nonzero (albeit small) ion flux, representing a nonambipolar
current for the prescribed input profiles.
In contrast, the local $\delta f$ model (\ref{eq:local}) is intrinsically ambipolar \cite{Kovrizhnykh,Rutherford},
meaning there is never a nonambipolar current
(to the order of accuracy of (\ref{eq:local})) for any choice of input profiles.
This property can be proved for the local $\delta f$ model by forming the
\begin{equation}
\sum_a \left< \int d^3v \frac{I\vpar}{\Omega_a} \left( ... \right) \right>
\label{eq:momentumConservation}
\end{equation}
moment of (\ref{eq:local}), using the momentum-conserving property of collisions, and integrating by parts in $\theta$.
It follows that the radial current must vanish to this order:
\begin{equation}
\sum_a \left< \int d^3v \, f_a \vda\cdot\nabla\psi\right> = 0
\label{eq:ambipolarity}
\end{equation}
where $S_a$ may be neglected.
However, when the operation
(\ref{eq:momentumConservation}) is applied to the global $\delta f$ equation (\ref{eq15}),
the new term arising from $\vdao\cdot\nabla g_a$ can balance the radial current in (\ref{eq:ambipolarity}).
The nonambipolar current reflects the fact that the input electric field profile is not self-consistent.
The situation is thus similar to neoclassical calculations for stellarators, and it may be possible to
iterate and adjust the input profiles to determine a radial electric field that results in zero radial
current. We do not attempt this calculation here. Such an electric field profile would not be meaningful unless
$B_{\theta}/B \ll 1$, since otherwise finite-gyroradius effects neglected in this drift-kinetic description would become
important\cite{Felix2010APS}. The resulting electric field profile would also only be meaningful if other sources of nonambipolar
particle transport \cite{Felix2010,DorfEr} are unimportant.

We close by mentioning another connection to stellarators.
It is common in stellarator neoclassical calculations to retain
poloidal $\ExB$ precession but to neglect radial derivatives of the non-Maxwellian
part of the distribution function\cite{Beidler}.
While this approach avoids the need to couple flux surfaces in the calculations,
it leads to non-conservation of energy and magnetic moment,
as discussed in \cite{meMonoenergetic} and the appendix of Ref. \cite{finiteErOmnigenous}.
Using PERFECT, which retains $\ExB$ precession, we are now able to directly compare
results with and without the radial derivative term, at least for axisymmetric plasmas.
As can be seen in the blue dashed curves in
figure (\ref{fig:PetersProfiles}).D-E,
neglect of the radial derivative term may lead to large errors in output quantities.

\begin{acknowledgments}
This work was supported by the US Department of Energy through grants DE-FG02-91ER-54109 and DE-FG02-93ER-54197.
Some of the computer simulations presented here
were carried out on the MIT PSFC parallel AMD Opteron/Infiniband cluster Loki.
Other simulations
used resources of the National Energy Research Scientific Computing Center (NERSC),
which is supported by the Office of Science of the U.S. Department of Energy under Contract No. DE-AC02-05CH11231.
M. L. was supported by the
Fusion Energy Postdoctoral Research Program
administered by the Oak Ridge Institute for Science and Education.
I. P. was supported by the International Postdoc grant of Vetenskapsr{\aa}det.
\changed{Conversations about this work with Grigory Kagan of Los Alamos National Laboratory and Mikhail Dorf of Lawrence Livermore National Laboratory are gratefully acknowledged.}
\end{acknowledgments}

\appendix

\section{Normalizations and equations implemented in the code PERFECT}

The quantities which must be supplied to solve (\ref{eq15}) are $m_a$, $Z_a$, $T_a\left( \psi \right)$,
$n_a\left( \psi \right)$, $\Phi_0 \left( \psi \right)$, $B\left( {\psi
,\theta } \right)$, $I\left( \psi \right)$, and the coordinate Jacobian
$J=\vect{B}\cdot \nabla \theta $. Any $\theta $ coordinate may be used:
$\arctan \left( {Z/\left[ {R-R_0 } \right]} \right)$, the angle
$\theta $ used to define Miller equilibrium\cite{Miller}, the poloidal Boozer angle, or any other poloidal angle.

The radial coordinate used in PERFECT is the normalized
flux $\psi_N =\psi /\psia $ where $\psia $ is the flux at the last
closed flux surface. The input quantities are specified in terms of some
dimensions $\Bar{{T}}$ (e.g. eV), $\Bar{{n}}$ (e.g. 10$^{20}$/m$^{3})$,
$\Bar{{\Phi }}$ (e.g. kV), $\Bar{{B}}$ (e.g. T), $\Bar{{R}}$ (e.g. m),
and $\bar{m}$ (e.g. the deuteron mass).
In other words, the quantities which are actually supplied as input to the
code are
$Z_a$,
$\hat{m}_a = m_a/\bar{m}$,
$\Hat{{T}}_a=T_a/\Bar{{T}}$,
$\Hat{{n}}_a=n_a/\Bar{{n}}$,
$\Hat{{\Phi }}=\Phi_0 /\Bar{{\Phi }}$,
$\Hat{{B}}=B/\Bar{{B}}$,
$\Hat{{I}}=I/\left( {\Bar{{R}}\Bar{{B}}} \right)$,
$\psiaHat =\psia /\left( {\Bar{{R}}^2\Bar{{B}}} \right)$,
and
$\Hat{{J}}=(\vect{B}\cdot \nabla \theta )\Bar{{R}}/\Bar{{B}}$.
We define
$\bar{v}=\sqrt{2\bar{T}/\bar{m}}$,
$\Delta =\bar{m}c\Bar{{v }}/(e\Bar{{B}}\Bar{{R}})$,
and
$\omega =c\Bar{{\Phi }}/(\Bar{{v }}\Bar{{R}}\Bar{{B}})$,
and the unknown part of the distribution function is normalized using
$g_a=\Delta\Bar{{n}}\Bar{{v }}^{-3}\Hat{{g}}_a$ where we will solve for $\hat{g}_a$.
We also define the collision frequency at the reference parameters
\begin{equation}
\label{eq97}
\bar{\nu}
=\frac{4\sqrt {2\pi }
e^4\Bar{{n}}\ln \Lambda }{3\sqrt{\bar{m}} \Bar{{T}}^{3/2}}
\end{equation}
and the associated normalized collisionality $\nu_r = \bar{\nu}\bar{R}/\bar{v}$.
Applying these normalizations,
and changing to $(\psi_N,\theta,x_a,\xi)$ coordinates,
the kinetic equation (\ref{eq15}) becomes (after some algebra)
\begin{equation}
\label{eq100}
 \Dot{{\psi }}_{Na} \frac{\partial \Hat{{g}}_a}{\partial \psi_N }+\Dot{{\theta
}}_a\frac{\partial \Hat{{g}}_a}{\partial \theta }+\Dot{{\xi }}_a\frac{\partial
\Hat{{g}}_a}{\partial \xi }+\Dot{{x}}_a\frac{\partial \Hat{{g}}_a}{\partial
x_a}
-{\nu_r }
\sum_b \Hat{{C}}_{ab}
-\hat{S}_a
= (1+\xi^2) D_a
\end{equation}
where
\begin{equation}
\label{eq53}
\Dot{{\psi }}_{Na} =-\Delta \frac{\sqrt{\hat{m}_a} \Hat{{T}}_a\Hat{{J}}\Hat{{I}}}{Z_a \psiaHat
\Hat{{B}}^3}x_a^2\frac{\left( {1+\xi^2} \right)}{2}\frac{\partial
\Hat{{B}}}{\partial \theta },
\end{equation}
\begin{equation}
\label{eq60}
 \Dot{{\theta }}_a=\frac{\Hat{{J}}x_a \xi \sqrt {\Hat{{T}}_a} }{\Hat{{B}}}
 +\frac{\omega\sqrt{\hat{m}_a}\Hat{{J}}\Hat{{I}}}{\psiaHat \Hat{{B}}^2}\frac{d\Hat{{\Phi
}}}{d\psi_N }
+\Delta x_a^2\frac{\left( {1+\xi^2}
\right)}{2}\frac{\sqrt{\hat{m}_a}\Hat{{T}}_a\Hat{{I}}\Hat{{J}}}{Z_a\Hat{{B}}^3\psiaHat
}\frac{\partial \Hat{{B}}}{\partial \psi_N }
-\Delta x_a^2\xi^2\frac{\sqrt{m}_a\Hat{{T}}_a\Hat{{J}}}{Z_a\psiaHat
\Hat{{B}}^2}\frac{d\Hat{{I}}}{d\psi_N },
\end{equation}
\begin{equation}
\label{eq64}
\Dot{{x}}_a
=\left( { \frac{\Delta x_a^3}{2Z_a}\frac{d\Hat{{T}}_a}{d\psi_N }
+\omega x_a\frac{d\Hat{{\Phi }}}{d\psi_N }}
\right)
\frac{\sqrt{\hat{m}_a}\Hat{{J}}\Hat{{I}}}{\psiaHat \Hat{{B}}^3}\frac{\left(
{1+\xi^2} \right)}{2}\frac{\partial \Hat{{B}}}{\partial \theta },
\end{equation}
\begin{equation}
\label{eq67}
 \Dot{{\xi }}_a=-\frac{\Hat{{J}}x_a\sqrt {\Hat{{T}}_a} }{2\Hat{{B}}^2}\left(
{1-\xi^2} \right)\frac{\partial \Hat{{B}}}{\partial \theta }\
+\frac{\omega\sqrt{\hat{m}_a}\Hat{{J}}\Hat{{I}}}{\psiaHat
\Hat{{B}}^3}\frac{\xi \left( {1-\xi^2} \right)}{2}
\frac{d\Hat{{\Phi }}}{d\psi_N }\frac{\partial
\Hat{{B}}}{\partial \theta }
 +\frac{\Delta x_a^2\sqrt{\hat{m}_a} \Hat{{T}}_a\Hat{{J}}}{2Z_a \psiaHat \Hat{{B}}^3}\xi \left( {1-\xi^2} \right)
\frac{d\Hat{{I}}}{d\psi_N }
\frac{\partial \Hat{{B}}}{\partial \theta } ,
\end{equation}
\begin{equation}
\hat{C}_{ab} = \frac{\sqrt{\hat{m}_a}}{\bar{\nu}}
\left(
C_{ab}\left\{\hat{g}_a,f_{\mathrm{m}b}\right\}
+
C_{ab}\left\{f_{\mathrm{m}a},\hat{g}_b\right\}
\right)
\end{equation}
is a normalized collision operator,
$C_{ab}$ is the Fokker-Planck operator,
$\Hat{{S}}_a=\Bar{{v }}^2\Bar{{R}}(\Delta \Bar{{n}})^{-1}\sqrt{\hat{m}_a}S_a$
is the normalized source,
and
\begin{equation}
D_a =
\frac{\hat{m}_a^2 \hat{n}_a \hat{I} \hat{J}}
{2\pi^{3/2} Z_a \sqrt{\hat{T}_a} \psiaHat \hat{B}^3}
\frac{\partial \hat{B}}{\partial \theta}
x_a^2 e^{-x_a^2}
\left( \frac{1}{\hat{n}_a} \frac{d\hat{n}_a}{d\psi_N}
+\frac{2Z_a \omega}{\Delta \hat{T}_a} \frac{d\hat{\Phi}}{d\psi_N}
+\left[x_a^2-\frac{3}{2}\right] \frac{1}{\hat{T}_a}\frac{d\hat{T}_a}{d\psi_N}
\right).
\end{equation}
Defining normalized perturbed Rosenbluth potentials $\Hat{{H}}_b$ and
$\Hat{{G}}_b$  by
$H_b=\Delta \Bar{{n}}\Bar{{v }}^{-3}v_b^2 \Hat{{H}}_b$
and
$G_b=\Delta \Bar{{n}}\Bar{{v }}^{-3}v_b^4 \Hat{{G}}_b$,
then the normalized collision operator may be written
$\Hat{{C}}_{ab}=
\Hat{{C}}_{ab}^L
+\Hat{{C}}_{ab}^E
+\Hat{{C}}_{ab}^D
+\Hat{{C}}_{ab}^H
+\Hat{{C}}_{ab}^G
$
where
\begin{equation}
\label{eq88}
\Hat{{C}}_{ab}^L
=\frac{\Hat{{\nu }}_{ab}^D}{2}
\frac{\partial }{\partial \xi }
\left( {1-\xi^2} \right)
\frac{\partial\Hat{{g}}_a}{\partial \xi },
\end{equation}
$\Hat{\nu }_{ab}^D =(3\sqrt \pi/4) \hat{n}_b Z_a^2 Z_b^2 \hat{T}_1^{-3/2} \left[ \erf\left( x_b
\right)-\Psi \left( x_b \right) \right]x_a^{-3}$,
$\Psi(x_b) =\left[ {\erf\left( x_b
\right)-2 \pi^{-1/2}x_b e^{-x_b^2}} \right]/(2x_b^2)$,
\begin{eqnarray}
\label{eq90}
\Hat{{C}}_{ab}^E
&=&
\frac{3\sqrt \pi \hat{n}_b Z_a^2 Z_b^2}
{4\hat{T}_a^{3/2}}
\left[
\frac{\Psi(x_b)}{x_a} \frac{\partial^2 \hat{g}_a}{\partial x_a^2}
+\frac{4}{\sqrt{\pi}}
\sqrt{\frac{\hat{m}_b}{\hat{m}_a}}
\left( \frac{\hat{T}_a}{\hat{T}_b}\right)^{3/2}
e^{-x_b^2} \hat{g}_a \right. \\
& & \left.
+\left\{
-2\frac{\hat{T}_a \hat{m}_b}{\hat{T}_b \hat{m}_a} \Psi(x_b)
\left( 1 - \frac{\hat{m}_a}{\hat{m}_b}\right)
+\frac{1}{x_a^2}
\left[
\erf(x_b^2) - \Psi(x_b)\right]
\right\}
\frac{\partial \hat{g}_a}{\partial x_a} \right] \nonumber
\end{eqnarray}
is the energy scattering operator,
\begin{equation}
\hat{C}_{ab}^D = 3\frac{\hat{n}_a Z_a^2 Z_b^2 \hat{m}_a}{\hat{T}_a^{3/2} \hat{m}_b}
e^{-x_a^2} \hat{g}_b,
\end{equation}
\begin{equation}
\hat{C}_{ab}^H = -\frac{3\hat{n}_a Z_a^2 Z_b^2 \hat{T}_b \hat{m}_a}{2\pi \hat{T}_a^{5/2} \hat{m}_b}
e^{-x_a^2}
\left[\left(1-\frac{\hat{m}_a}{\hat{m}_b}\right)
x_b\frac{\partial \hat{H}_b}{\partial x_b}
+\hat{H}_b\right],
\end{equation}
and
\begin{equation}
\hat{C}_{ab}^G = \frac{3\hat{n}_a Z_a^2 Z_b^2 \hat{T}_b \hat{m}_a}{2\pi \hat{T}_a^{5/2} \hat{m}_b}
e^{-x_a^2} x_a^2\frac{\partial^2 \hat{G}_b}{\partial x_b^2},
\end{equation}
and the potentials are determined by the following two Poisson-like equations:
\begin{equation}
\label{eq94}
\left[ {\frac{\partial }{\partial x_b}x_b^2\frac{\partial }{\partial
x_b}+\frac{\partial }{\partial \xi }\left( {1-\xi^2} \right)\frac{\partial
}{\partial \xi }} \right]\Hat{{H}}_b=-4\pi x_b^2\Hat{{g}}_b
\end{equation}
and
\begin{equation}
\label{eq95}
\left[ {\frac{\partial }{\partial x_b}x_b^2\frac{\partial }{\partial
x_b}+\frac{\partial }{\partial \xi }\left( {1-\xi^2} \right)\frac{\partial
}{\partial \xi }} \right]\Hat{{G}}_b=2x_b^2\Hat{{H}}_b.
\end{equation}

Next,
the problem is cast into Legendre modes, writing the distribution function
as
\begin{equation}
\label{eq101}
\Hat{{g}}_a=\sum\limits_{\ell =0}^{N_\xi-1} {\hat{g}_{a,\ell} \left( {\psi_N ,\theta ,x_a}
\right)P_\ell \left( \xi \right)}
\end{equation}
where $N_\xi$ is the number of Legendre polynomials retained,
with analogous expansions for $\Hat{{H}}_b$ and $\Hat{{G}}_b$. The operation
$(2L+1)2^{-1}\int_{-1}^1 {d\xi \;} P_L \left( \xi
\right)\left( {\mbox{\thinspace \thinspace \thinspace \thinspace }\cdot
\mbox{\thinspace \thinspace \thinspace \thinspace }} \right)$
is applied to the kinetic equation.
To evaluate the integrals of the collisionless terms,
the following Legendre identities may be used:
\begin{eqnarray}
\label{eq108}
 \frac{2L+1}{2}\int_{-1}^1 {d\xi \mbox{\thinspace }} \xi P_L \left( \xi
\right)P_\ell \left( \xi \right)
& =&\frac{L+1}{2L+3}\delta_{L+1,\ell } +\frac{L}{2L-1}\delta_{L-1,\ell } ,
\end{eqnarray}
\begin{eqnarray}
\label{eq110}
 \frac{2L+1}{2}\int_{-1}^1 {d\xi } \left( {1+\xi^2}
\right)P_L \left( \xi \right)P_\ell \left( \xi \right)
&=&\frac{2\left(
{3L^2+3L-2} \right)}{\left( {2L+3} \right)\left( {2L-1} \right)}\delta
_{L,\ell }
 +\frac{L-1}{2L-3}\frac{L}{2L-1}\delta_{L-2,\ell } \nonumber \\
&&+\frac{L+2}{2L+5}\frac{L+1}{2L+3}\delta_{L+2,\ell } ,
\end{eqnarray}
\begin{equation}
\label{eq112}
\frac{2L+1}{2}\int_{-1}^1 {d\xi } \left( {1-\xi^2}
\right)P_L \left( \xi \right)\frac{dP_\ell }{d\xi }=\frac{\left( {L+1}
\right)\left( {L+2} \right)}{2L+3}\delta_{L+1,\ell } -\frac{\left( {L-1}
\right)L}{2L-1}\delta_{L-1,\ell },
\end{equation}
and
\begin{eqnarray}
\label{eq113}
 \frac{2L+1}{2}\int_{-1}^1 {d\xi } \left( {1-\xi^2}
\right)\xi P_L \left( \xi \right)\frac{dP_\ell }{d\xi }
&=&\frac{\left( {L+1}
\right)L}{\left( {2L-1} \right)\left( {2L+3} \right)}\delta_{L,\ell }
 +\frac{\left( {L+3} \right)\left( {L+2} \right)\left( {L+1} \right)}{\left(
{2L+5} \right)\left( {2L+3} \right)}\delta_{L+2,\ell } \nonumber \\
&&-\frac{L\left( {L-1}
\right)\left( {L-2} \right)}{\left( {2L-3} \right)\left( {2L-1}
\right)}\delta_{L-2,\ell },
\end{eqnarray}
where $\delta_{x,y} $ is a Kronecker delta.
As a result, the kinetic equation (\ref{eq100}) takes the form
\begin{equation}
\label{eq104}
\sum\limits_{\ell =0}^{N_\xi-1} {M_{a,L,\ell } } g_{a,\ell}
-\delta_{L,0}\hat{S}_a
=\left( \frac{4}{3}\delta_{L,0} + \frac{2}{3} \delta_{L,2}\right) D_a
\end{equation}
where
\begin{equation}
\label{eq114}
M_{a,L,\ell } =\Dot{{\psi }}_{a,L,\ell } \frac{\partial }{\partial \psi_N
}+\Dot{{\theta }}_{a,L,\ell } \frac{\partial }{\partial \theta
}+\Dot{{x}}_{a,L,\ell } \frac{\partial }{\partial x_a}+M_{a,L,\ell }^{\left( \xi
\right)} -{\nu_r } \sum_b \Hat{{C}}_{ab} \delta_{L,\ell},
\end{equation}
\begin{eqnarray}
\label{eq115}
\Dot{{\psi }}_{a,L,\ell }
&=&-
\frac{\Delta\sqrt{\hat{m}_a} \Hat{{T}}_a\Hat{{J}}\Hat{{I}}}{2 Z_a\psiaHat
\Hat{{B}}^3}x_a^2\frac{\partial \Hat{{B}}}{\partial \theta }\left[
 \frac{2\left( {3L^2+3L-2} \right)}{\left( {2L+3} \right)\left( {2L-1}
\right)}\delta_{L,\ell } \right.  \\
&&\left.
\hspace{1in} +\frac{L-1}{2L-3}\frac{L}{2L-1}\delta_{L-2,\ell }
+\frac{L+2}{2L+5}\frac{L+1}{2L+3}\delta_{L+2,\ell }
 \right],\nonumber
\end{eqnarray}

\begin{eqnarray}
\label{eq116}
 \Dot{{\theta }}_{a,L,\ell }
 &=&\frac{\Hat{{J}}x_a\sqrt {\Hat{{T}}_a}
}{\Hat{{B}}}\left( {\frac{L+1}{2L+3}\delta_{L+1,\ell }
+\frac{L}{2L-1}\delta_{L-1,\ell } } \right)
 + \frac{\omega \sqrt{\hat{m}_a}\Hat{{J}}\Hat{{I}}}{\psiaHat
\Hat{{B}}^2}\frac{d\Hat{{\Phi }}}{d\psi_N }\delta_{L,\ell }  \\
&& +\Delta x_a^2\frac{\sqrt{\hat{m}_a}\Hat{{T}}_a\Hat{{J}}}{Z_a \Hat{{B}}^2\psiaHat
}\frac{1}{\left( {2L+3} \right)\left( {2L-1} \right)}\left[ {\left(
{3L^2+3L-2} \right)\frac{\Hat{{I}}}{\Hat{{B}}}\frac{\partial
\Hat{{B}}}{\partial \psi_N }-\left( {2L^2+2L-1}
\right)\frac{d\Hat{{I}}}{d\psi_N }} \right]\delta_{L,\ell } \nonumber \\
&& +\Delta x_a^2\frac{\sqrt{\hat{m}_a}\Hat{{T}}_a\Hat{{J}}}{Z_a\Hat{{B}}^2\psiaHat }\left(
{\frac{\Hat{{I}}}{2\Hat{{B}}}\frac{\partial \Hat{{B}}}{\partial \psi_N
}-\frac{d\Hat{{I}}}{d\psi_N }} \right)\left(
{\frac{L-1}{2L-3}\frac{L}{2L-1}\delta_{L-2,\ell }
+\frac{L+2}{2L+5}\frac{L+1}{2L+3}\delta_{L+2,\ell } } \right) ,\nonumber
\end{eqnarray}

\begin{eqnarray}
\label{eq117}
\Dot{{x}}_{a,L,\ell }
&=&\left( { \frac{\Delta x_a^3}{2 Z_a}\frac{d\Hat{{T}}_a}{d\psi_N}
+\omega x_a\frac{d\Hat{{\Phi }}}{d\psi_N }}
\right)\frac{\sqrt{\hat{m}_a}\Hat{{J}}\Hat{{I}}}{2\psiaHat \Hat{{B}}^3}\frac{\partial
\Hat{{B}}}{\partial \theta }
\left[
 \frac{2\left( {3L^2+3L-2} \right)}{\left( {2L+3} \right)\left( {2L-1}
\right)}\delta_{L,\ell } \right. \nonumber \\
&&\hspace{1in} \left. +\frac{L-1}{2L-3}\frac{L}{2L-1}\delta_{L-2,\ell }
+\frac{L+2}{2L+5}\frac{L+1}{2L+3}\delta_{L+2,\ell }
  \right],
\end{eqnarray}
and, from (\ref{eq112}) and (\ref{eq113}),
\begin{eqnarray}
\label{eq118}
 M_{a,L,\ell }^{\left( \xi \right)}
 &=&-\frac{\Hat{{J}}x_a\sqrt {\Hat{{T}}_a}
}{2\Hat{{B}}^2}\frac{\partial \Hat{{B}}}{\partial \theta }\left[
{\frac{\left( {L+1} \right)\left( {L+2} \right)}{2L+3}\delta_{L+1,\ell }
-\frac{\left( {L-1} \right)L}{2L-1}\delta_{L-1,\ell } } \right] \nonumber \\
&& +\left( {\omega \frac{d\Hat{{\Phi }}}{d\psi_N }\Hat{{I}}
+\frac{\Delta x_a^2\Hat{{T}}_a}{Z_a}
\frac{d\Hat{{I}}}{d\psi_N }}
\right)\frac{\sqrt{\hat{m}_a}\Hat{{J}}}{2\psiaHat \Hat{{B}}^3}\frac{\partial
\Hat{{B}}}{\partial \theta } \\
&&\times \left[
 \frac{\left( {L+1} \right)L}{\left( {2L-1} \right)\left( {2L+3}
\right)}\delta_{L,\ell }
 +\frac{\left( {L+3} \right)\left( {L+2} \right)\left( {L+1} \right)}{\left(
{2L+5} \right)\left( {2L+3} \right)}\delta_{L+2,\ell } -\frac{L\left( {L-1}
\right)\left( {L-2} \right)}{\left( {2L-3} \right)\left( {2L-1}
\right)}\delta_{L-2,\ell }
 \right]. \nonumber
\end{eqnarray}
The collision operator is diagonal in the Legendre representation, with
expressions (\ref{eq88}) and (\ref{eq94})-(\ref{eq95}) simplifying using
$(\partial /\partial \xi)\left( {1-\xi^2} \right)\partial P_L /\partial\xi=-L\left( {L+1} \right)P_L \left( \xi
\right)$.

Once the kinetic equation is solved, the parallel flow is computed from
\begin{equation}
\label{eq133}
V_{||a}
= \frac{1}{n_a} \int d^3v\, \vpar g_a
= \frac{4\pi \Delta \bar{v}\hat{T}_a^2}
{3 \hat{n}_a \hat{m}_a^2}
\int _0^{\infty} dx_a \, x_a^3 \hat{g}_{a,1}.
\end{equation}
The particle flux is
\begin{equation}
\label{eq143}
\Gamma_a ={V}'\left\langle {\int {d^3v \mbox{\thinspace }}
g_a\vma \cdot \nabla \psi_N } \right\rangle
=\frac{\Bar{{n}}\Bar{{v }}}{\Bar{{B}}}\frac{\pi \Delta^2}{\psiaHat }\Hat{{\Gamma }}_a
\end{equation}
where
\begin{equation}
\label{eq142}
\Hat{{\Gamma }}_a=-\frac{\hat{m}_a \hat{I}}{Z_a} \left(\frac{\Hat{{T}}_a}{\hat{m}_a}\right)^{5/2}
\int_0^{2\pi } {d\theta
\frac{1}{\Hat{{B}}^3}\frac{\partial \Hat{{B}}}{\partial \theta }}
\int_0^\infty {dx_a} \left( {\frac{8}{3}\Hat{{g}}_{a,0}
+\frac{4}{15}\Hat{{g}}_{a,2} } \right)x_a^4.
\end{equation}
The momentum flux is
\begin{equation}
\label{eq149}
\Pi_a ={V}'\left\langle {\int {d^3v \mbox{\thinspace }}
g_a \frac{Iv_{\vert \vert } }{B}\vma \cdot \nabla \psi_N }
\right\rangle
=-\frac{\Bar{{v
}}^2\Bar{{n}}\Bar{{R}}}{\Bar{{B}}}\frac{\Delta^2\pi }{\psiaHat
}\Hat{{\Pi }}_a
\end{equation}
where
\begin{equation}
\label{eq148}
\Hat{{\Pi }}_a=-\frac{\hat{m}_a \hat{I}^2}{Z_a}
\left(\frac{\Hat{{T}}_a}{\hat{m}_a}\right)^3
\int_0^{2\pi } {d\theta
\frac{1}{\Hat{{B}}^4}\frac{\partial \Hat{{B}}}{\partial \theta }}
\int_0^\infty {dx_a} \left( {\frac{16}{15}\Hat{{g}}_{a,1}
+\frac{4}{35}\Hat{{g}}_{a,3} } \right)x_a^5.
\end{equation}
The heat flux is
\begin{equation}
\label{eq154}
Q_a={V}'\left\langle {\int {d^3v \mbox{\thinspace }} g_a\frac{m_a v
^2}{2}\vma \cdot \nabla \psi_N } \right\rangle
=\frac{\bar{m}\Bar{{n}}\Bar{{v }}^3}{\Bar{{B}}}\frac{\pi \Delta
^2}{2\psiaHat }\Hat{{Q}}_a,
\end{equation}
where
\begin{equation}
\label{eq153}
\Hat{{Q}}_a=
-\frac{\hat{m}_a \hat{T}_a \hat{I}}{Z_a}
\left(\frac{\Hat{{T}}_a}{\hat{m}_a}\right)^{5/2}
\int_0^{2\pi } {d\theta
\frac{1}{\Hat{{B}}^3}\frac{\partial \Hat{{B}}}{\partial \theta }}
\int_0^\infty {dx_a} \left( {\frac{8}{3}\Hat{{g}}_{a,0}
+\frac{4}{15}\Hat{{g}}_{a,2} } \right)x_a^6.
\end{equation}

\bibliography{PedestalCodePaper}

\begin{thebibliography}{62}
\expandafter\ifx\csname natexlab\endcsname\relax\def\natexlab#1{#1}\fi
\expandafter\ifx\csname bibnamefont\endcsname\relax
  \def\bibnamefont#1{#1}\fi
\expandafter\ifx\csname bibfnamefont\endcsname\relax
  \def\bibfnamefont#1{#1}\fi
\expandafter\ifx\csname citenamefont\endcsname\relax
  \def\citenamefont#1{#1}\fi
\expandafter\ifx\csname url\endcsname\relax
  \def\url#1{\texttt{#1}}\fi
\expandafter\ifx\csname urlprefix\endcsname\relax\def\urlprefix{URL }\fi
\providecommand{\bibinfo}[2]{#2}
\providecommand{\eprint}[2][]{\url{#2}}

\bibitem[{\citenamefont{Hinton and Hazeltine}(1976)}]{HintonHazeltine}
\bibinfo{author}{\bibfnamefont{F.~L.} \bibnamefont{Hinton}} \bibnamefont{and}
  \bibinfo{author}{\bibfnamefont{R.~D.} \bibnamefont{Hazeltine}},
  \bibinfo{journal}{Rev. Mod. Phys.} \textbf{\bibinfo{volume}{48}},
  \bibinfo{pages}{239} (\bibinfo{year}{1976}).

\bibitem[{\citenamefont{Helander and Sigmar}(2002)}]{PerBook}
\bibinfo{author}{\bibfnamefont{P.}~\bibnamefont{Helander}} \bibnamefont{and}
  \bibinfo{author}{\bibfnamefont{D.~J.} \bibnamefont{Sigmar}},
  \emph{\bibinfo{title}{Collisional Transport in Magnetized Plasmas}}
  (\bibinfo{publisher}{Cambridge University Press},
  \bibinfo{address}{Cambridge}, \bibinfo{year}{2002}).

\bibitem[{\citenamefont{Lin et~al.}(1995)\citenamefont{Lin, Tang, and
  Lee}}]{Lin1}
\bibinfo{author}{\bibfnamefont{Z.}~\bibnamefont{Lin}},
  \bibinfo{author}{\bibfnamefont{W.~M.} \bibnamefont{Tang}}, \bibnamefont{and}
  \bibinfo{author}{\bibfnamefont{W.~W.} \bibnamefont{Lee}},
  \bibinfo{journal}{Phys. Plasmas} \textbf{\bibinfo{volume}{2}},
  \bibinfo{pages}{2975} (\bibinfo{year}{1995}).

\bibitem[{\citenamefont{Lin et~al.}(1997)\citenamefont{Lin, Tang, and
  Lee}}]{Lin2}
\bibinfo{author}{\bibfnamefont{Z.}~\bibnamefont{Lin}},
  \bibinfo{author}{\bibfnamefont{W.~M.} \bibnamefont{Tang}}, \bibnamefont{and}
  \bibinfo{author}{\bibfnamefont{W.~W.} \bibnamefont{Lee}},
  \bibinfo{journal}{Phys. Rev. Lett.} \textbf{\bibinfo{volume}{78}},
  \bibinfo{pages}{456} (\bibinfo{year}{1997}).

\bibitem[{\citenamefont{Wang et~al.}(2001)\citenamefont{Wang, Hinton, and
  Wong}}]{Wang1}
\bibinfo{author}{\bibfnamefont{W.~X.} \bibnamefont{Wang}},
  \bibinfo{author}{\bibfnamefont{F.~L.} \bibnamefont{Hinton}},
  \bibnamefont{and} \bibinfo{author}{\bibfnamefont{S.~K.} \bibnamefont{Wong}},
  \bibinfo{journal}{Phys. Rev. Lett.} \textbf{\bibinfo{volume}{87}},
  \bibinfo{pages}{055002} (\bibinfo{year}{2001}).

\bibitem[{\citenamefont{Chang et~al.}(2004)\citenamefont{Chang, Ku, and
  Weitzner}}]{CSChang1}
\bibinfo{author}{\bibfnamefont{C.~S.} \bibnamefont{Chang}},
  \bibinfo{author}{\bibfnamefont{S.}~\bibnamefont{Ku}}, \bibnamefont{and}
  \bibinfo{author}{\bibfnamefont{H.}~\bibnamefont{Weitzner}},
  \bibinfo{journal}{Phys. Plasmas} \textbf{\bibinfo{volume}{11}},
  \bibinfo{pages}{2649} (\bibinfo{year}{2004}).

\bibitem[{\citenamefont{Wang et~al.}(2004)\citenamefont{Wang, Tang, Hinton,
  Zakharov, White, and Manickam}}]{Wang2}
\bibinfo{author}{\bibfnamefont{W.~X.} \bibnamefont{Wang}},
  \bibinfo{author}{\bibfnamefont{W.~M.} \bibnamefont{Tang}},
  \bibinfo{author}{\bibfnamefont{F.~L.} \bibnamefont{Hinton}},
  \bibinfo{author}{\bibfnamefont{L.~E.} \bibnamefont{Zakharov}},
  \bibinfo{author}{\bibfnamefont{R.~B.} \bibnamefont{White}}, \bibnamefont{and}
  \bibinfo{author}{\bibfnamefont{J.}~\bibnamefont{Manickam}},
  \bibinfo{journal}{Comp. Phys. Comm.} \textbf{\bibinfo{volume}{164}},
  \bibinfo{pages}{178} (\bibinfo{year}{2004}).

\bibitem[{\citenamefont{Chang and Ku}(2006)}]{CSChang2}
\bibinfo{author}{\bibfnamefont{C.~S.} \bibnamefont{Chang}} \bibnamefont{and}
  \bibinfo{author}{\bibfnamefont{S.}~\bibnamefont{Ku}},
  \bibinfo{journal}{Contrib. Plasma Phys.} \textbf{\bibinfo{volume}{46}},
  \bibinfo{pages}{496} (\bibinfo{year}{2006}).

\bibitem[{\citenamefont{Wang et~al.}(2006)\citenamefont{Wang, Rewoldt, Tang,
  Hinton, Manickam, Zakharov, White, and Kaye}}]{Wang3}
\bibinfo{author}{\bibfnamefont{W.~X.} \bibnamefont{Wang}},
  \bibinfo{author}{\bibfnamefont{G.}~\bibnamefont{Rewoldt}},
  \bibinfo{author}{\bibfnamefont{W.~M.} \bibnamefont{Tang}},
  \bibinfo{author}{\bibfnamefont{F.~L.} \bibnamefont{Hinton}},
  \bibinfo{author}{\bibfnamefont{J.}~\bibnamefont{Manickam}},
  \bibinfo{author}{\bibfnamefont{L.~E.} \bibnamefont{Zakharov}},
  \bibinfo{author}{\bibfnamefont{R.~B.} \bibnamefont{White}}, \bibnamefont{and}
  \bibinfo{author}{\bibfnamefont{S.}~\bibnamefont{Kaye}},
  \bibinfo{journal}{Phys. Plasmas} \textbf{\bibinfo{volume}{13}},
  \bibinfo{pages}{082501} (\bibinfo{year}{2006}).

\bibitem[{\citenamefont{Kolesnikov et~al.}(2010)\citenamefont{Kolesnikov, Wang,
  Hinton, Rewoldt, and Tang}}]{Kolesnikov}
\bibinfo{author}{\bibfnamefont{R.~A.} \bibnamefont{Kolesnikov}},
  \bibinfo{author}{\bibfnamefont{W.~X.} \bibnamefont{Wang}},
  \bibinfo{author}{\bibfnamefont{F.~L.} \bibnamefont{Hinton}},
  \bibinfo{author}{\bibfnamefont{G.}~\bibnamefont{Rewoldt}}, \bibnamefont{and}
  \bibinfo{author}{\bibfnamefont{W.~M.} \bibnamefont{Tang}},
  \bibinfo{journal}{Phys. Plasmas} \textbf{\bibinfo{volume}{17}},
  \bibinfo{pages}{022506} (\bibinfo{year}{2010}).

\bibitem[{\citenamefont{Vernay et~al.}(2010)\citenamefont{Vernay, Brunner,
  Villard, McMillan, Jollier, Tran, Bottino, and Graves}}]{ORB5}
\bibinfo{author}{\bibfnamefont{T.}~\bibnamefont{Vernay}},
  \bibinfo{author}{\bibfnamefont{S.}~\bibnamefont{Brunner}},
  \bibinfo{author}{\bibfnamefont{L.}~\bibnamefont{Villard}},
  \bibinfo{author}{\bibfnamefont{B.~F.} \bibnamefont{McMillan}},
  \bibinfo{author}{\bibfnamefont{S.}~\bibnamefont{Jollier}},
  \bibinfo{author}{\bibfnamefont{T.~M.} \bibnamefont{Tran}},
  \bibinfo{author}{\bibfnamefont{A.}~\bibnamefont{Bottino}}, \bibnamefont{and}
  \bibinfo{author}{\bibfnamefont{J.~P.} \bibnamefont{Graves}},
  \bibinfo{journal}{Phys. Plasmas} \textbf{\bibinfo{volume}{17}},
  \bibinfo{pages}{122301} (\bibinfo{year}{2010}).

\bibitem[{\citenamefont{Xu et~al.}(2007)\citenamefont{Xu, Xiong, Dorr,
  Hittinger, Bodi, Candy, Cohen, Cohen, Colella, Kerbel et~al.}}]{Xu1}
\bibinfo{author}{\bibfnamefont{X.~Q.} \bibnamefont{Xu}},
  \bibinfo{author}{\bibfnamefont{Z.}~\bibnamefont{Xiong}},
  \bibinfo{author}{\bibfnamefont{M.~R.} \bibnamefont{Dorr}},
  \bibinfo{author}{\bibfnamefont{J.~A.} \bibnamefont{Hittinger}},
  \bibinfo{author}{\bibfnamefont{K.}~\bibnamefont{Bodi}},
  \bibinfo{author}{\bibfnamefont{J.}~\bibnamefont{Candy}},
  \bibinfo{author}{\bibfnamefont{B.~I.} \bibnamefont{Cohen}},
  \bibinfo{author}{\bibfnamefont{R.~H.} \bibnamefont{Cohen}},
  \bibinfo{author}{\bibfnamefont{P.}~\bibnamefont{Colella}},
  \bibinfo{author}{\bibfnamefont{G.~D.} \bibnamefont{Kerbel}},
  \bibnamefont{et~al.}, \bibinfo{journal}{Nucl. Fusion}
  \textbf{\bibinfo{volume}{47}}, \bibinfo{pages}{809} (\bibinfo{year}{2007}).

\bibitem[{\citenamefont{Xu}(2008)}]{Xu2}
\bibinfo{author}{\bibfnamefont{X.~Q.} \bibnamefont{Xu}},
  \bibinfo{journal}{Phys. Rev. E} \textbf{\bibinfo{volume}{78}},
  \bibinfo{pages}{016406} (\bibinfo{year}{2008}).

\bibitem[{\citenamefont{Koh et~al.}(2012)\citenamefont{Koh, Chang, Ku, Menard,
  Weitzner, and Choe}}]{XGCBootstrap}
\bibinfo{author}{\bibfnamefont{S.}~\bibnamefont{Koh}},
  \bibinfo{author}{\bibfnamefont{C.~S.} \bibnamefont{Chang}},
  \bibinfo{author}{\bibfnamefont{S.}~\bibnamefont{Ku}},
  \bibinfo{author}{\bibfnamefont{J.~E.} \bibnamefont{Menard}},
  \bibinfo{author}{\bibfnamefont{H.}~\bibnamefont{Weitzner}}, \bibnamefont{and}
  \bibinfo{author}{\bibfnamefont{W.}~\bibnamefont{Choe}},
  \bibinfo{journal}{Phys. Plasmas} \textbf{\bibinfo{volume}{19}},
  \bibinfo{pages}{072505} (\bibinfo{year}{2012}).

\bibitem[{\citenamefont{Dorf et~al.}(2013{\natexlab{a}})\citenamefont{Dorf,
  Cohen, Dorr, Rognlien, Hittinger, Compton, Colella, Martin, and
  McCorquodale}}]{Dorf}
\bibinfo{author}{\bibfnamefont{M.~A.} \bibnamefont{Dorf}},
  \bibinfo{author}{\bibfnamefont{R.~H.} \bibnamefont{Cohen}},
  \bibinfo{author}{\bibfnamefont{M.}~\bibnamefont{Dorr}},
  \bibinfo{author}{\bibfnamefont{T.}~\bibnamefont{Rognlien}},
  \bibinfo{author}{\bibfnamefont{J.}~\bibnamefont{Hittinger}},
  \bibinfo{author}{\bibfnamefont{J.}~\bibnamefont{Compton}},
  \bibinfo{author}{\bibfnamefont{P.}~\bibnamefont{Colella}},
  \bibinfo{author}{\bibfnamefont{D.}~\bibnamefont{Martin}}, \bibnamefont{and}
  \bibinfo{author}{\bibfnamefont{P.}~\bibnamefont{McCorquodale}},
  \bibinfo{journal}{Phys. Plasmas} \textbf{\bibinfo{volume}{20}},
  \bibinfo{pages}{012513} (\bibinfo{year}{2013}{\natexlab{a}}).

\bibitem[{\citenamefont{Wong and Chan}(2011)}]{WongChan}
\bibinfo{author}{\bibfnamefont{S.~K.} \bibnamefont{Wong}} \bibnamefont{and}
  \bibinfo{author}{\bibfnamefont{V.~S.} \bibnamefont{Chan}},
  \bibinfo{journal}{Plasma Phys. Controlled Fusion}
  \textbf{\bibinfo{volume}{53}}, \bibinfo{pages}{095005}
  (\bibinfo{year}{2011}).

\bibitem[{\citenamefont{Belli and Candy}(2012)}]{NEOFP}
\bibinfo{author}{\bibfnamefont{E.~A.} \bibnamefont{Belli}} \bibnamefont{and}
  \bibinfo{author}{\bibfnamefont{J.}~\bibnamefont{Candy}},
  \bibinfo{journal}{Plasma Phys. Controlled Fusion}
  \textbf{\bibinfo{volume}{54}}, \bibinfo{pages}{015015}
  (\bibinfo{year}{2012}).

\bibitem[{\citenamefont{Parker and Catto}(2012)}]{JeffParker}
\bibinfo{author}{\bibfnamefont{J.~B.} \bibnamefont{Parker}} \bibnamefont{and}
  \bibinfo{author}{\bibfnamefont{P.~J.} \bibnamefont{Catto}},
  \bibinfo{journal}{Plasma Phys. Controlled Fusion}
  \textbf{\bibinfo{volume}{54}}, \bibinfo{pages}{085011}
  (\bibinfo{year}{2012}).

\bibitem[{\citenamefont{Kagan and Catto}(2008)}]{Grisha1}
\bibinfo{author}{\bibfnamefont{G.}~\bibnamefont{Kagan}} \bibnamefont{and}
  \bibinfo{author}{\bibfnamefont{P.~J.} \bibnamefont{Catto}},
  \bibinfo{journal}{Plasma Phys. Controlled Fusion}
  \textbf{\bibinfo{volume}{50}}, \bibinfo{pages}{085010}
  (\bibinfo{year}{2008}).

\bibitem[{\citenamefont{Landreman and Ernst}(2012)}]{usPedestal}
\bibinfo{author}{\bibfnamefont{M.}~\bibnamefont{Landreman}} \bibnamefont{and}
  \bibinfo{author}{\bibfnamefont{D.~R.} \bibnamefont{Ernst}},
  \bibinfo{journal}{Plasma Phys. Controlled Fusion}
  \textbf{\bibinfo{volume}{54}}, \bibinfo{pages}{115006}
  (\bibinfo{year}{2012}).

\bibitem[{\citenamefont{{R. J. Groebner and T. H.
  Osborne}}(1998)}]{GroebnerOsborne}
\bibinfo{author}{\bibnamefont{{R. J. Groebner and T. H. Osborne}}},
  \bibinfo{journal}{Phys. Plasmas} \textbf{\bibinfo{volume}{5}},
  \bibinfo{pages}{1800} (\bibinfo{year}{1998}), \bibinfo{note}{fig. 2.}

\bibitem[{\citenamefont{{C.F. Maggi, R.J. Groebner, N. Oyama, R. Sartori, L.D.
  Horton, A.C.C. Sips,W. Suttrop and the ASDEX Upgrade Team, A. Leonard, T.C.
  Luce, M.R.Wade and the DIII-D Team, et al}}(2007)}]{Maggi}
\bibinfo{author}{\bibnamefont{{C.F. Maggi, R.J. Groebner, N. Oyama, R. Sartori,
  L.D. Horton, A.C.C. Sips,W. Suttrop and the ASDEX Upgrade Team, A. Leonard,
  T.C. Luce, M.R.Wade and the DIII-D Team, et al}}}, \bibinfo{journal}{Nucl.
  Fusion} \textbf{\bibinfo{volume}{47}}, \bibinfo{pages}{535}
  (\bibinfo{year}{2007}).

\bibitem[{\citenamefont{Corre et~al.}(2008)\citenamefont{Corre, Joffrin,
  Monier-Garbet, Andrew, Arnoux, Beurskens, Brezinsek, Brix, Buttery, and
  I~Coffey}}]{Corre}
\bibinfo{author}{\bibfnamefont{Y.}~\bibnamefont{Corre}},
  \bibinfo{author}{\bibfnamefont{E.}~\bibnamefont{Joffrin}},
  \bibinfo{author}{\bibfnamefont{P.}~\bibnamefont{Monier-Garbet}},
  \bibinfo{author}{\bibfnamefont{Y.}~\bibnamefont{Andrew}},
  \bibinfo{author}{\bibfnamefont{G.}~\bibnamefont{Arnoux}},
  \bibinfo{author}{\bibfnamefont{M.}~\bibnamefont{Beurskens}},
  \bibinfo{author}{\bibfnamefont{S.}~\bibnamefont{Brezinsek}},
  \bibinfo{author}{\bibfnamefont{M.}~\bibnamefont{Brix}},
  \bibinfo{author}{\bibfnamefont{R.}~\bibnamefont{Buttery}}, \bibnamefont{and}
  \bibinfo{author}{\bibfnamefont{e.~a.} \bibnamefont{I~Coffey}},
  \bibinfo{journal}{Plasma Phys. Controlled Fusion}
  \textbf{\bibinfo{volume}{50}}, \bibinfo{pages}{115012}
  (\bibinfo{year}{2008}).

\bibitem[{\citenamefont{Groebner et~al.}(2009)\citenamefont{Groebner, Osborne,
  Leonard, and Fenstermacher}}]{Groebner}
\bibinfo{author}{\bibfnamefont{R.~J.} \bibnamefont{Groebner}},
  \bibinfo{author}{\bibfnamefont{T.~H.} \bibnamefont{Osborne}},
  \bibinfo{author}{\bibfnamefont{A.~W.} \bibnamefont{Leonard}},
  \bibnamefont{and} \bibinfo{author}{\bibfnamefont{M.~E.}
  \bibnamefont{Fenstermacher}}, \bibinfo{journal}{Nucl. Fusion}
  \textbf{\bibinfo{volume}{49}}, \bibinfo{pages}{045013}
  (\bibinfo{year}{2009}).

\bibitem[{\citenamefont{{T. W. Morgan, H. Meyer, D. Temple, and G. J.
  Tallents}}(2010)}]{Morgan}
\bibinfo{author}{\bibnamefont{{T. W. Morgan, H. Meyer, D. Temple, and G. J.
  Tallents}}}, in \emph{\bibinfo{booktitle}{37th EPS Conf. Plasma Phys.}}
  (\bibinfo{address}{Dublin}, \bibinfo{year}{2010}), p.
  \bibinfo{pages}{P5.122}.

\bibitem[{\citenamefont{Diallo et~al.}(2011)\citenamefont{Diallo, Maingi,
  Kubota, Sontag, Osborne, Podesta, Bell, LeBlanc, Menard, and
  Sabbagh}}]{Diallo}
\bibinfo{author}{\bibfnamefont{A.}~\bibnamefont{Diallo}},
  \bibinfo{author}{\bibfnamefont{R.}~\bibnamefont{Maingi}},
  \bibinfo{author}{\bibfnamefont{S.}~\bibnamefont{Kubota}},
  \bibinfo{author}{\bibfnamefont{A.}~\bibnamefont{Sontag}},
  \bibinfo{author}{\bibfnamefont{T.}~\bibnamefont{Osborne}},
  \bibinfo{author}{\bibfnamefont{M.}~\bibnamefont{Podesta}},
  \bibinfo{author}{\bibfnamefont{R.~E.} \bibnamefont{Bell}},
  \bibinfo{author}{\bibfnamefont{B.~P.} \bibnamefont{LeBlanc}},
  \bibinfo{author}{\bibfnamefont{J.}~\bibnamefont{Menard}}, \bibnamefont{and}
  \bibinfo{author}{\bibfnamefont{S.}~\bibnamefont{Sabbagh}},
  \bibinfo{journal}{Nucl. Fusion} \textbf{\bibinfo{volume}{51}},
  \bibinfo{pages}{103031} (\bibinfo{year}{2011}).

\bibitem[{\citenamefont{{H. Meyer, M. F. M. De Bock, N. J. Conway, S. J.
  Freethy, K. Gibson, J. Hiratsuka, A. Kirk, C. A. Michael, T. Morgan, R.
  Scannell, et al}}(2011)}]{Meyer}
\bibinfo{author}{\bibnamefont{{H. Meyer, M. F. M. De Bock, N. J. Conway, S. J.
  Freethy, K. Gibson, J. Hiratsuka, A. Kirk, C. A. Michael, T. Morgan, R.
  Scannell, et al}}}, \bibinfo{journal}{Nucl. Fusion}
  \textbf{\bibinfo{volume}{51}}, \bibinfo{pages}{113011}
  (\bibinfo{year}{2011}).

\bibitem[{\citenamefont{{A.C. Sontag, J.M. Canik, R. Maingi, J. Manickam, P.B.
  Snyder, R.E. Bell, S.P. Gerhardt, S. Kubota, B.P. LeBlanc, D. Mueller, T.H.
  Osborne and K.L. Tritz}}(2011)}]{Sontag}
\bibinfo{author}{\bibnamefont{{A.C. Sontag, J.M. Canik, R. Maingi, J. Manickam,
  P.B. Snyder, R.E. Bell, S.P. Gerhardt, S. Kubota, B.P. LeBlanc, D. Mueller,
  T.H. Osborne and K.L. Tritz}}}, \bibinfo{journal}{Nucl. Fusion}
  \textbf{\bibinfo{volume}{51}}, \bibinfo{pages}{103022}
  (\bibinfo{year}{2011}).

\bibitem[{\citenamefont{{P. Sauter, T. Putterich, F. Ryter, E. Viezzer, E.
  Wolfrum, G.D. Conway, R. Fischer, B. Kurzan, R.M. McDermott, S.K. Rathgeber
  and the ASDEX Upgrade Team}}(2012)}]{SauterProfiles}
\bibinfo{author}{\bibnamefont{{P. Sauter, T. Putterich, F. Ryter, E. Viezzer,
  E. Wolfrum, G.D. Conway, R. Fischer, B. Kurzan, R.M. McDermott, S.K.
  Rathgeber and the ASDEX Upgrade Team}}}, \bibinfo{journal}{Nucl. Fusion}
  \textbf{\bibinfo{volume}{52}}, \bibinfo{pages}{012001}
  (\bibinfo{year}{2012}).

\bibitem[{\citenamefont{{R. Maingi, D.P. Boyle, J.M. Canik, S.M. Kaye, C.H.
  Skinner, J.P. Allain, M.G. Bell, R.E. Bell, S.P. Gerhardt, T.K. Gray, et
  al}}(2012)}]{Maingi}
\bibinfo{author}{\bibnamefont{{R. Maingi, D.P. Boyle, J.M. Canik, S.M. Kaye,
  C.H. Skinner, J.P. Allain, M.G. Bell, R.E. Bell, S.P. Gerhardt, T.K. Gray, et
  al}}}, \bibinfo{journal}{Nucl. Fusion} \textbf{\bibinfo{volume}{52}},
  \bibinfo{pages}{083001} (\bibinfo{year}{2012}).

\bibitem[{\citenamefont{{T. P\"utterich, E. Viezzer, R. Dux, R. M. McDermott,
  and the ASDEX Upgrade team}}(2012)}]{Putterich}
\bibinfo{author}{\bibnamefont{{T. P\"utterich, E. Viezzer, R. Dux, R. M.
  McDermott, and the ASDEX Upgrade team}}}, \bibinfo{journal}{Nucl. Fusion}
  \textbf{\bibinfo{volume}{52}}, \bibinfo{pages}{083013}
  (\bibinfo{year}{2012}).

\bibitem[{\citenamefont{Lapillonne et~al.}(2010)\citenamefont{Lapillonne,
  McMillan, Gorler, Brunner, Dannert, Jenko, Merz, and Villard}}]{GENESource}
\bibinfo{author}{\bibfnamefont{X.}~\bibnamefont{Lapillonne}},
  \bibinfo{author}{\bibfnamefont{B.~F.} \bibnamefont{McMillan}},
  \bibinfo{author}{\bibfnamefont{T.}~\bibnamefont{Gorler}},
  \bibinfo{author}{\bibfnamefont{S.}~\bibnamefont{Brunner}},
  \bibinfo{author}{\bibfnamefont{T.}~\bibnamefont{Dannert}},
  \bibinfo{author}{\bibfnamefont{F.}~\bibnamefont{Jenko}},
  \bibinfo{author}{\bibfnamefont{F.}~\bibnamefont{Merz}}, \bibnamefont{and}
  \bibinfo{author}{\bibfnamefont{L.}~\bibnamefont{Villard}},
  \bibinfo{journal}{Phys. Plasmas} \textbf{\bibinfo{volume}{17}},
  \bibinfo{pages}{112321} (\bibinfo{year}{2010}).

\bibitem[{\citenamefont{Kagan and Catto}(2010{\natexlab{a}})}]{GrishaNeo}
\bibinfo{author}{\bibfnamefont{G.}~\bibnamefont{Kagan}} \bibnamefont{and}
  \bibinfo{author}{\bibfnamefont{P.~J.} \bibnamefont{Catto}},
  \bibinfo{journal}{Plasma Phys. Controlled Fusion}
  \textbf{\bibinfo{volume}{52}}, \bibinfo{pages}{055004}
  (\bibinfo{year}{2010}{\natexlab{a}}).

\bibitem[{\citenamefont{Kagan and
  Catto}(2010{\natexlab{b}})}]{GrishaNeoErratum}
\bibinfo{author}{\bibfnamefont{G.}~\bibnamefont{Kagan}} \bibnamefont{and}
  \bibinfo{author}{\bibfnamefont{P.~J.} \bibnamefont{Catto}},
  \bibinfo{journal}{Plasma Phys. Controlled Fusion}
  \textbf{\bibinfo{volume}{52}}, \bibinfo{pages}{079801}
  (\bibinfo{year}{2010}{\natexlab{b}}).

\bibitem[{\citenamefont{Kagan and Catto}(2010{\natexlab{c}})}]{GrishaPRL}
\bibinfo{author}{\bibfnamefont{G.}~\bibnamefont{Kagan}} \bibnamefont{and}
  \bibinfo{author}{\bibfnamefont{P.~J.} \bibnamefont{Catto}},
  \bibinfo{journal}{Phys. Rev. Lett.} \textbf{\bibinfo{volume}{105}},
  \bibinfo{pages}{045002} (\bibinfo{year}{2010}{\natexlab{c}}).

\bibitem[{\citenamefont{Pusztai and Catto}(2010{\natexlab{a}})}]{Istvan}
\bibinfo{author}{\bibfnamefont{I.}~\bibnamefont{Pusztai}} \bibnamefont{and}
  \bibinfo{author}{\bibfnamefont{P.~J.} \bibnamefont{Catto}},
  \bibinfo{journal}{Plasma Phys. Controlled Fusion}
  \textbf{\bibinfo{volume}{52}}, \bibinfo{pages}{075016}
  (\bibinfo{year}{2010}{\natexlab{a}}).

\bibitem[{\citenamefont{Pusztai and Catto}(2010{\natexlab{b}})}]{IstvanErratum}
\bibinfo{author}{\bibfnamefont{I.}~\bibnamefont{Pusztai}} \bibnamefont{and}
  \bibinfo{author}{\bibfnamefont{P.~J.} \bibnamefont{Catto}},
  \bibinfo{journal}{Plasma Phys. Controlled Fusion}
  \textbf{\bibinfo{volume}{52}}, \bibinfo{pages}{119801}
  (\bibinfo{year}{2010}{\natexlab{b}}).

\bibitem[{\citenamefont{Catto et~al.}(2013)\citenamefont{Catto, Parra, Kagan,
  Parker, Pusztai, and Landreman}}]{Peter}
\bibinfo{author}{\bibfnamefont{P.~J.} \bibnamefont{Catto}},
  \bibinfo{author}{\bibfnamefont{F.~I.} \bibnamefont{Parra}},
  \bibinfo{author}{\bibfnamefont{G.}~\bibnamefont{Kagan}},
  \bibinfo{author}{\bibfnamefont{J.~B.} \bibnamefont{Parker}},
  \bibinfo{author}{\bibfnamefont{I.}~\bibnamefont{Pusztai}}, \bibnamefont{and}
  \bibinfo{author}{\bibfnamefont{M.}~\bibnamefont{Landreman}},
  \bibinfo{journal}{Plasma Phys. Controlled Fusion}
  \textbf{\bibinfo{volume}{55}}, \bibinfo{pages}{045009}
  (\bibinfo{year}{2013}).

\bibitem[{\citenamefont{Parra and Catto}(2010{\natexlab{a}})}]{Felix2010}
\bibinfo{author}{\bibfnamefont{F.~I.} \bibnamefont{Parra}} \bibnamefont{and}
  \bibinfo{author}{\bibfnamefont{P.~J.} \bibnamefont{Catto}},
  \bibinfo{journal}{Plasma Phys. Controlled Fusion}
  \textbf{\bibinfo{volume}{52}}, \bibinfo{pages}{045004}
  (\bibinfo{year}{2010}{\natexlab{a}}).

\bibitem[{\citenamefont{Hazeltine}(1973)}]{Hazeltine}
\bibinfo{author}{\bibfnamefont{R.~D.} \bibnamefont{Hazeltine}},
  \bibinfo{journal}{Plasma Phys.} \textbf{\bibinfo{volume}{15}},
  \bibinfo{pages}{77} (\bibinfo{year}{1973}).

\bibitem[{\citenamefont{Boozer}(1980)}]{Boozer80}
\bibinfo{author}{\bibfnamefont{A.~H.} \bibnamefont{Boozer}},
  \bibinfo{journal}{Phys. Fluids} \textbf{\bibinfo{volume}{23}},
  \bibinfo{pages}{904} (\bibinfo{year}{1980}).

\bibitem[{\citenamefont{Parra and Catto}(2008)}]{Parra}
\bibinfo{author}{\bibfnamefont{F.~I.} \bibnamefont{Parra}} \bibnamefont{and}
  \bibinfo{author}{\bibfnamefont{P.~J.} \bibnamefont{Catto}},
  \bibinfo{journal}{Plasma Phys. Controlled Fusion}
  \textbf{\bibinfo{volume}{50}}, \bibinfo{pages}{065014}
  (\bibinfo{year}{2008}).

\bibitem[{\citenamefont{Waltz et~al.}(2002)\citenamefont{Waltz, Candy, and
  Rosenbluth}}]{GYROSources}
\bibinfo{author}{\bibfnamefont{R.~E.} \bibnamefont{Waltz}},
  \bibinfo{author}{\bibfnamefont{J.~M.} \bibnamefont{Candy}}, \bibnamefont{and}
  \bibinfo{author}{\bibfnamefont{M.~N.} \bibnamefont{Rosenbluth}},
  \bibinfo{journal}{Phys. Plasmas} \textbf{\bibinfo{volume}{9}},
  \bibinfo{pages}{1938} (\bibinfo{year}{2002}).

\bibitem[{\citenamefont{Gorler et~al.}(2011)\citenamefont{Gorler, Lapillonne,
  Brunner, Dannert, Jenko, Merz, and Told}}]{globalGENE}
\bibinfo{author}{\bibfnamefont{T.}~\bibnamefont{Gorler}},
  \bibinfo{author}{\bibfnamefont{X.}~\bibnamefont{Lapillonne}},
  \bibinfo{author}{\bibfnamefont{S.}~\bibnamefont{Brunner}},
  \bibinfo{author}{\bibfnamefont{T.}~\bibnamefont{Dannert}},
  \bibinfo{author}{\bibfnamefont{F.}~\bibnamefont{Jenko}},
  \bibinfo{author}{\bibfnamefont{F.}~\bibnamefont{Merz}}, \bibnamefont{and}
  \bibinfo{author}{\bibfnamefont{D.}~\bibnamefont{Told}}, \bibinfo{journal}{J.
  Comp. Phys.} \textbf{\bibinfo{volume}{230}}, \bibinfo{pages}{7053}
  (\bibinfo{year}{2011}).

\bibitem[{\citenamefont{Helander}(2000)}]{PerPotato}
\bibinfo{author}{\bibfnamefont{P.}~\bibnamefont{Helander}},
  \bibinfo{journal}{Phys. Plasmas} \textbf{\bibinfo{volume}{7}},
  \bibinfo{pages}{2878} (\bibinfo{year}{2000}).

\bibitem[{\citenamefont{Trefethen}(2000)}]{Trefethen}
\bibinfo{author}{\bibfnamefont{L.~N.} \bibnamefont{Trefethen}},
  \emph{\bibinfo{title}{Spectral methods in Matlab}}
  (\bibinfo{publisher}{SIAM}, \bibinfo{address}{Philadelphia},
  \bibinfo{year}{2000}).

\bibitem[{\citenamefont{Landreman and Ernst}(2013)}]{speedGrids}
\bibinfo{author}{\bibfnamefont{M.}~\bibnamefont{Landreman}} \bibnamefont{and}
  \bibinfo{author}{\bibfnamefont{D.~R.} \bibnamefont{Ernst}},
  \bibinfo{journal}{J. Comp. Phys.} \textbf{\bibinfo{volume}{243}},
  \bibinfo{pages}{130} (\bibinfo{year}{2013}).

\bibitem[{\citenamefont{Amestoy et~al.}(2001)\citenamefont{Amestoy, Duff,
  Koster, and L'Excellent}}]{MUMPS:1}
\bibinfo{author}{\bibfnamefont{P.~R.} \bibnamefont{Amestoy}},
  \bibinfo{author}{\bibfnamefont{I.~S.} \bibnamefont{Duff}},
  \bibinfo{author}{\bibfnamefont{J.}~\bibnamefont{Koster}}, \bibnamefont{and}
  \bibinfo{author}{\bibfnamefont{J.-Y.} \bibnamefont{L'Excellent}},
  \bibinfo{journal}{SIAM Journal on Matrix Analysis and Applications}
  \textbf{\bibinfo{volume}{23}}, \bibinfo{pages}{15} (\bibinfo{year}{2001}).

\bibitem[{\citenamefont{Li et~al.}(1999)\citenamefont{Li, Demmel, Gilbert,
  Grigori, Shao, and Yamazaki}}]{superlu1}
\bibinfo{author}{\bibfnamefont{X.}~\bibnamefont{Li}},
  \bibinfo{author}{\bibfnamefont{J.}~\bibnamefont{Demmel}},
  \bibinfo{author}{\bibfnamefont{J.}~\bibnamefont{Gilbert}},
  \bibinfo{author}{\bibfnamefont{L.}~\bibnamefont{Grigori}},
  \bibinfo{author}{\bibfnamefont{M.}~\bibnamefont{Shao}}, \bibnamefont{and}
  \bibinfo{author}{\bibfnamefont{I.}~\bibnamefont{Yamazaki}},
  \bibinfo{type}{Tech. Rep.} \bibinfo{number}{LBNL-44289},
  \bibinfo{institution}{Lawrence Berkeley National Laboratory}
  (\bibinfo{year}{1999}),
  \bibinfo{note}{\url{http://crd.lbl.gov/~xiaoye/SuperLU/}. Last update: August
  2011}.

\bibitem[{\citenamefont{Li and Demmel}(2003)}]{superlu2}
\bibinfo{author}{\bibfnamefont{X.~S.} \bibnamefont{Li}} \bibnamefont{and}
  \bibinfo{author}{\bibfnamefont{J.~W.} \bibnamefont{Demmel}},
  \bibinfo{journal}{ACM Trans. Mathematical Software}
  \textbf{\bibinfo{volume}{29}}, \bibinfo{pages}{110} (\bibinfo{year}{2003}).

\bibitem[{\citenamefont{Saad and Schultz}(1986)}]{GMRES}
\bibinfo{author}{\bibfnamefont{Y.}~\bibnamefont{Saad}} \bibnamefont{and}
  \bibinfo{author}{\bibfnamefont{M.~H.} \bibnamefont{Schultz}},
  \bibinfo{journal}{{SIAM J. Sci. and Stat. Comput.}}
  \textbf{\bibinfo{volume}{7}}, \bibinfo{pages}{856} (\bibinfo{year}{1986}).

\bibitem[{\citenamefont{Sleijpen and Fokkema}(1993)}]{BICGstabl}
\bibinfo{author}{\bibfnamefont{G.~L.~G.} \bibnamefont{Sleijpen}}
  \bibnamefont{and} \bibinfo{author}{\bibfnamefont{D.~R.}
  \bibnamefont{Fokkema}}, \bibinfo{journal}{{Electr. Trans. Num. Anal.}}
  \textbf{\bibinfo{volume}{1}}, \bibinfo{pages}{11} (\bibinfo{year}{1993}).

\bibitem[{\citenamefont{Balay et~al.}(Accessed October 6,
  2012)\citenamefont{Balay, Brown, Buschelman, Gropp, Kaushik, Knepley,
  McInnes, Smith, and Zhang}}]{petsc-web-page}
\bibinfo{author}{\bibfnamefont{S.}~\bibnamefont{Balay}},
  \bibinfo{author}{\bibfnamefont{J.}~\bibnamefont{Brown}},
  \bibinfo{author}{\bibfnamefont{K.}~\bibnamefont{Buschelman}},
  \bibinfo{author}{\bibfnamefont{W.~D.} \bibnamefont{Gropp}},
  \bibinfo{author}{\bibfnamefont{D.}~\bibnamefont{Kaushik}},
  \bibinfo{author}{\bibfnamefont{M.~G.} \bibnamefont{Knepley}},
  \bibinfo{author}{\bibfnamefont{L.~C.} \bibnamefont{McInnes}},
  \bibinfo{author}{\bibfnamefont{B.~F.} \bibnamefont{Smith}}, \bibnamefont{and}
  \bibinfo{author}{\bibfnamefont{H.}~\bibnamefont{Zhang}},
  \emph{\bibinfo{title}{{PETSc} {W}eb page}} (\bibinfo{year}{Accessed October
  6, 2012}), \bibinfo{note}{{http://www.mcs.anl.gov/petsc}}.

\bibitem[{\citenamefont{Balay et~al.}(2012)\citenamefont{Balay, Brown, ,
  Buschelman, Eijkhout, Gropp, Kaushik, Knepley, McInnes, Smith
  et~al.}}]{petsc-user-ref}
\bibinfo{author}{\bibfnamefont{S.}~\bibnamefont{Balay}},
  \bibinfo{author}{\bibfnamefont{J.}~\bibnamefont{Brown}}, ,
  \bibinfo{author}{\bibfnamefont{K.}~\bibnamefont{Buschelman}},
  \bibinfo{author}{\bibfnamefont{V.}~\bibnamefont{Eijkhout}},
  \bibinfo{author}{\bibfnamefont{W.~D.} \bibnamefont{Gropp}},
  \bibinfo{author}{\bibfnamefont{D.}~\bibnamefont{Kaushik}},
  \bibinfo{author}{\bibfnamefont{M.~G.} \bibnamefont{Knepley}},
  \bibinfo{author}{\bibfnamefont{L.~C.} \bibnamefont{McInnes}},
  \bibinfo{author}{\bibfnamefont{B.~F.} \bibnamefont{Smith}},
  \bibnamefont{et~al.}, \bibinfo{type}{Tech. Rep.} \bibinfo{number}{ANL-95/11 -
  Revision 3.3}, \bibinfo{institution}{Argonne National Laboratory}
  (\bibinfo{year}{2012}).

\bibitem[{\citenamefont{Miller et~al.}(1998)\citenamefont{Miller, Chu, Greene,
  Lin-Liu, and Waltz}}]{Miller}
\bibinfo{author}{\bibfnamefont{R.~L.} \bibnamefont{Miller}},
  \bibinfo{author}{\bibfnamefont{M.~S.} \bibnamefont{Chu}},
  \bibinfo{author}{\bibfnamefont{J.~M.} \bibnamefont{Greene}},
  \bibinfo{author}{\bibfnamefont{Y.~R.} \bibnamefont{Lin-Liu}},
  \bibnamefont{and} \bibinfo{author}{\bibfnamefont{R.~E.} \bibnamefont{Waltz}},
  \bibinfo{journal}{Phys. Plasmas} \textbf{\bibinfo{volume}{5}},
  \bibinfo{pages}{973} (\bibinfo{year}{1998}).

\bibitem[{\citenamefont{Kovrizhnykh}(1969)}]{Kovrizhnykh}
\bibinfo{author}{\bibfnamefont{L.~M.} \bibnamefont{Kovrizhnykh}},
  \bibinfo{journal}{Sov. Phys. JETP} \textbf{\bibinfo{volume}{29}},
  \bibinfo{pages}{475} (\bibinfo{year}{1969}).

\bibitem[{\citenamefont{Rutherford}(1970)}]{Rutherford}
\bibinfo{author}{\bibfnamefont{P.~H.} \bibnamefont{Rutherford}},
  \bibinfo{journal}{Phys. Fluids} \textbf{\bibinfo{volume}{13}},
  \bibinfo{pages}{482} (\bibinfo{year}{1970}).

\bibitem[{\citenamefont{Parra and Catto}(2010{\natexlab{b}})}]{Felix2010APS}
\bibinfo{author}{\bibfnamefont{F.~I.} \bibnamefont{Parra}} \bibnamefont{and}
  \bibinfo{author}{\bibfnamefont{P.~J.} \bibnamefont{Catto}},
  \bibinfo{journal}{Phys. Plasmas} \textbf{\bibinfo{volume}{17}},
  \bibinfo{pages}{056106} (\bibinfo{year}{2010}{\natexlab{b}}).

\bibitem[{\citenamefont{Dorf et~al.}(2013{\natexlab{b}})\citenamefont{Dorf,
  Cohen, Simakov, and Joseph}}]{DorfEr}
\bibinfo{author}{\bibfnamefont{M.}~\bibnamefont{Dorf}},
  \bibinfo{author}{\bibfnamefont{R.~H.} \bibnamefont{Cohen}},
  \bibinfo{author}{\bibfnamefont{A.~N.} \bibnamefont{Simakov}},
  \bibnamefont{and} \bibinfo{author}{\bibfnamefont{I.}~\bibnamefont{Joseph}},
  \bibinfo{journal}{Phys. Plasmas} \textbf{\bibinfo{volume}{20}},
  \bibinfo{pages}{082515} (\bibinfo{year}{2013}{\natexlab{b}}).

\bibitem[{\citenamefont{{Beidler, C. D., et al.}}(2007)}]{Beidler}
\bibinfo{author}{\bibnamefont{{Beidler, C. D., et al.}}},
  \bibinfo{journal}{Proceedings of the 17th {I}nternational {T}oki {C}onference
  and 16th {I}nternational {S}tellarator/{H}eliotron {W}orkshop, {T}oki}
  (\bibinfo{year}{2007}).

\bibitem[{\citenamefont{Landreman}(2011)}]{meMonoenergetic}
\bibinfo{author}{\bibfnamefont{M.}~\bibnamefont{Landreman}},
  \bibinfo{journal}{Plasma Phys. Controlled Fusion}
  \textbf{\bibinfo{volume}{53}}, \bibinfo{pages}{082003}
  (\bibinfo{year}{2011}).

\bibitem[{\citenamefont{Landreman and Catto}(2013)}]{finiteErOmnigenous}
\bibinfo{author}{\bibfnamefont{M.}~\bibnamefont{Landreman}} \bibnamefont{and}
  \bibinfo{author}{\bibfnamefont{P.~J.} \bibnamefont{Catto}},
  \bibinfo{journal}{Plasma Phys. Controlled Fusion}
  \textbf{\bibinfo{volume}{55}}, \bibinfo{pages}{095017}
  (\bibinfo{year}{2013}).

\end{thebibliography}

\end{document}